\documentclass[reprint,notitlepage,tightenlines,nofootinbib,superscriptaddress]{revtex4-1}
\usepackage[utf8]{inputenc}
\usepackage{rotating}
\usepackage{amsfonts}
\usepackage{amsthm}
\usepackage{amssymb,hyperref, amsmath}
\usepackage{mathtools}
\usepackage{color}
\usepackage{graphicx}
\graphicspath{ {image/} }
\usepackage{multirow}
\usepackage{float}
\usepackage{subfigure} % use for side-by-side figures
\usepackage{soul}
\usepackage{hepunits}
\newcommand*\diff{\mathop{}\!\mathrm{d}}

\usepackage{braket}
\usepackage{bm}
\usepackage{txfonts}
\usepackage{mathrsfs} % in order to use mathscr
\definecolor{darkblue}{RGB}{68,60,150}
\begin{document}
 \bibliographystyle{apsrev4-1}
 \title{Frame dependence of transition form factors in light-front dynamics}
\author{Meijian Li}
\email{meijianl@iastate.edu}
\affiliation{Department of Physics and Astronomy, Iowa State University, Ames, IA, 50011}
\author{Yang Li}
\email{leeyoung@iastate.edu}
\affiliation{Department of Physics and Astronomy, Iowa State University, Ames, IA, 50011}
\affiliation{Hebei Key Laboratory of Compact Fusion, Langfang 065001, China }
\affiliation{ENN Science and Technology Development Co., Ltd., Langfang 065001, China}
\author{Pieter Maris}
\email{pmaris@iastate.edu}
\affiliation{Department of Physics and Astronomy, Iowa State University, Ames, IA, 50011}
\author{James P. Vary}
\email{jvary@iastate.edu}
\affiliation{Department of Physics and Astronomy, Iowa State University, Ames, IA, 50011}
\date{\today}
\begin{abstract}
We calculate the transition form factor between vector and pseudoscalar quarkonia in both the timelike and the spacelike region using light-front dynamics. We investigate the frame dependence of the form factors for heavy quarkonia with light-front wavefunctions calculated from the valence Fock sector. This dependence could serve as a measure for the Lorentz symmetry violation arising from the Fock-space truncation. We suggest using frames with minimal longitudinal momentum transfer for calculations in the valence Fock sector, namely the Drell-Yan frame for the space-like region and a specific longitudinal frame for the timelike region; at $q^2=0$ these two frames give the same result.
We also use the transition form factor in the timelike region to investigate the electromagnetic Dalitz decay $\psi_A\to \psi_B l^+l^-$ ($l = e,\mu$) and predict the effective mass spectrum of the lepton pair.
\end{abstract}

\maketitle
\section{Introduction}

The electromagnetic (EM) transition between quarkonium states,
which occurs via emission of a photon, $\psi_A\to \psi_B \gamma$, offers insights into the internal structure and the dynamics of such systems. The magnetic dipole (M1) transition, which takes place between pseudoscalar and vector mesons ($\psi_A, \psi_B = \mathcal V, \mathcal P$ or $\mathcal P, \mathcal V$), has been detected with strong signals~\cite{PDG2018} and stimulates various theoretical investigations~\cite{Brambilla:2005zw, Donald_JPsi, Bc_NR, cc_GI,Pineda:2013lta}. Similarly, the EM Dalitz decay~\cite{Dalitz:1951aj}, $\psi_A\to \psi_B l^+ l^-$, can be treated by coupling a virtual photon to the final lepton pair. The Dalitz decay, also known as the leptonic conversion decay, provides additional information about the meson structure owing to the virtual photon kinematics. Though widely observed in the light meson sector, such as $\phi\to\pi^0 e^+ e^-$~\cite{Anastasi:2016qga}, $\phi\to\eta e^+ e^-$~\cite{Achasov:2000ne,Babusci:2014ldz}, and $\omega\to\pi^0 e^+ e^-$~\cite{Akhmetshin:2005vy, Adlarson:2016hpp}, only a few such decays have been detected in the heavy sector. The observed Dalitz decays of quarkonium are decays to a light meson plus a lepton pair, such as $J/\psi\to \eta e^+ e^-$,  $J/\psi\to \eta' e^+ e^-$~\cite{Ablikim:2014nro}, and $\psi(3686)\to \eta' e^+ e^-$~\cite{Ablikim:2018xxs}. We investigate the M1 EM Dalitz decay with initial and final mesons both being heavy quarkonia, in the hope of providing another probe of the interaction of quarkonium states with photons. 
%({\color{red}Yang:Just curious: are there any available experimental data? In terms of what form? Can we compare with experiments?}{\color{blue}The experimental data is a distribution function of the lepton pair's invariant mass. They also present the transition form factor by fitting to the simple pole approximation. If our calculated transitions are measured, I think we can compare with experiments on the decay width (like our Fig.7) and the transition form factor (like our Fig.4.5).})

The roles of the underlying strong dynamics in those processes are encoded within the $q^2$-dependent transition form factor $V(q^2)$, which arises from the Lorentz structure decomposition of the hadron matrix element $\bra{\psi_B (P)} J^\mu(0) \ket{\psi_A(P'=P+q)}$. $q^2$ is the square of the momentum transfer between $\psi_A$ and $\psi_B$, and is also the square of the invariant mass of the lepton pair in the Dalitz decay.

\begin{figure}[htp]
  \subfigure[\ $n\to n$ \label{fig:nton}]
{\includegraphics[width=0.4\textwidth]{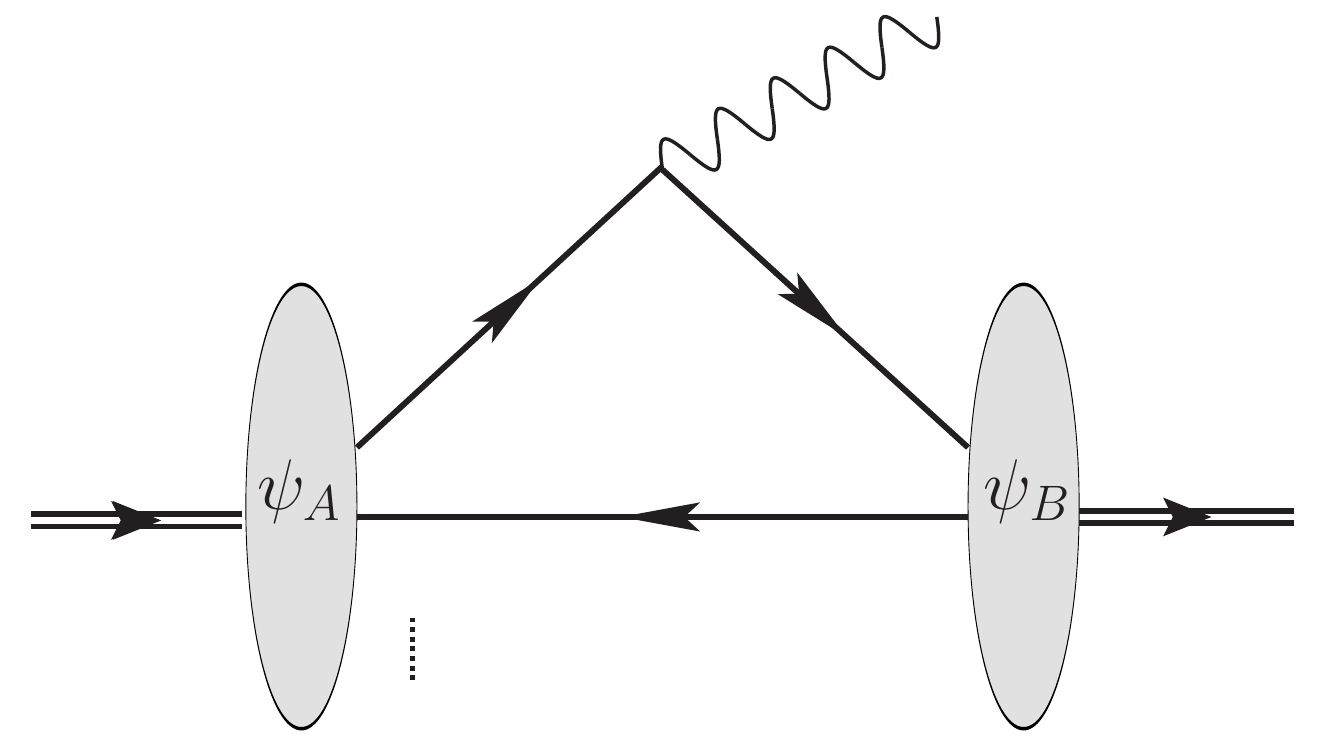}
} 
  \subfigure[\ $n+2 \to n$ \label{fig:np2ton}]
{\includegraphics[width=0.4\textwidth]{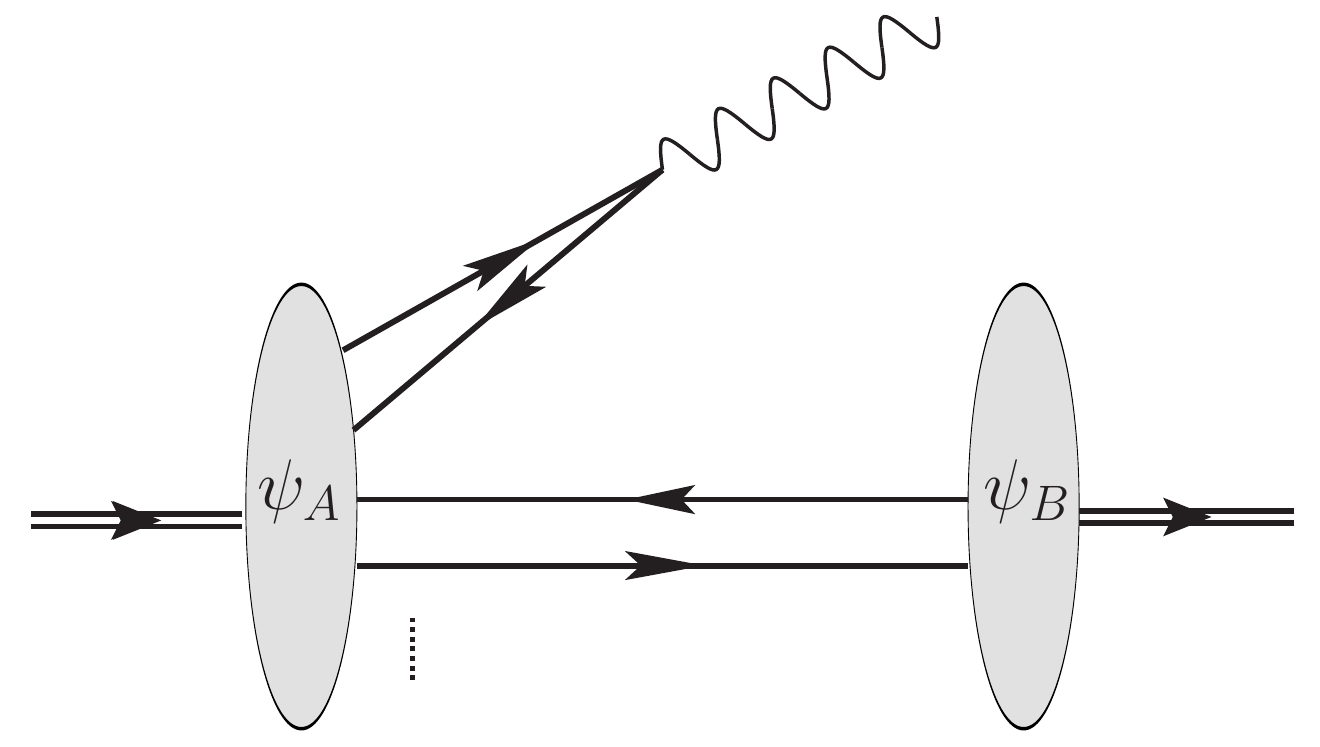}
}   
  \caption{\label{fig:partonN}The two dominant contributions to the transition $\psi_A\to\psi_B\gamma^{(*)}$ on the light front: (a) the parton-number-conserving term, and (b) the parton-number-non-conserving term. Diagrams are light-front time ordered. Light-front time flows from the left to the right.}
\end{figure}

The transition form factor is Lorentz invariant--it should not depend on the choice of the current components or the reference frames. 
In general, in light-front dynamics, the transition form factor receives two major contributions, a parton-number-conserving term where the photon couples to a parton, and a parton-number-non-conserving term where a quark-antiquark pair from the initial state annihilates into a photon, as illustrated in Fig.~\ref{fig:partonN}. 
Each diagram could contribute differently when a different current component is used or a different frame is chosen~\cite{Sawicki:1992qj,DEMELO1998574,BRODSKY1999239,Bakker:2003up}. In practical calculations the Fock space is truncated, so we only have access to part of the contributions. For instance, if the light-front wavefunctions of the mesons are solved in a truncation retaining only the valence Fock sector, contributions like Fig.~\ref{fig:np2ton} cannot be accessed directly\footnote{In principle, evaluation of observables in a truncated Fock space requires a renormalization of the operator. We omit such a renormalization in the present work.}. The transition form factor is evaluated from the available finite Fock space, and a dependence on the current components or the frames could arise. In such situations, knowing the current or frame dependence could help quantify theoretical uncertainties. One further issue to resolve is whether there exists a preferred current or a preferred frame such that the neglected contributions from higher Fock sectors could be minimized.

Dependence on current components when extracting the elastic and transition form factors on the light front has been studied extensively for
various systems and theories~\cite{DEMELO1998574,CARBONELL1998215,DEMELO1998574,BRODSKY1999239,
  PhysRevD.88.025036,PhysRevD.65.094043}. The $J^+$ current has gained the reputation as the ``good current'' for the simplicity in
evaluating the elastic form factor in the spacelike region in the Drell-Yan frame. The covariant formulation of light-front dynamics (CLFD) provides a procedure to unambiguously extract the transition form factors in the Drell-Yan frame (see Ref.~\cite{CARBONELL1998215} for a review). To explore some alternatives, we investigated the M1 transition form factor through different current components ($J^+$ and $J^\perp$) and different magnetic projections ($m_j=0,\pm 1$) of the vector meson~\cite{Li:2018uif}. There, we have shown that, at least in the context of heavy quarkonia with valence Fock sector light-front wavefunctions, using the transverse current $J^\perp$ in conjunction with the $m_j=0$ state of the vector meson is preferred to other choices. With the transverse current echoing the 3-dimensional current density operator, this choice employs the dominant spin components of the light-front wavefunctions, and connects well with the non-relativistic limit of the heavy system. While this preference applies to any choice of frame, the calculation of the final results was carried out in the Drell-Yan frame~\cite{Li:2018uif}.

It is the purpose of this work to further investigate the frame dependence of transition form factors. Studies in the literature have revealed that the elastic and transition form factors could have different results
when evaluated in different
reference frames~\cite{Cheng:1996if, BRODSKY1999239, Ji:2000rd, Bakker:2003up, Li:2017uug}.
 Such frame dependence is closely
related to the Fock-space truncation that omits the non-valence contributions. 
%({\color{red}The frame dependence of elastic form factors and transition form factors has been investigated in the literature [16-17, 23, 25]. Is this an accurate summary of these work?}
%{\color{blue} This is what I learned from those references, please let me know if you find it different.})
In this work, instead of looking into a few selective frames, we sample all frames with the same $q^2$. 

In particular, we sample various frames by allowing the transferred momentum $q^2$ to be apportioned between
the
longitudinal direction and the transverse direction.
%({\color{red}P.Maris: I'm not quite sure what you mean by this sentence.}{\color{blue}M.Li: I want to explain what is the difference among those frames that we define. I replaced ``in the sense that'' by ``where'' in the above sentence.})
 For example, in the Drell-Yan frame ($q^+=0$), all the transfered momentum is
in the transverse direction and none in the longitudinal, providing one limit to our frame selection. The other limit is referred
to as the longitudinal frame, where $q^2$ is solely in the longitudinal direction. 
We decompose $q^2$ into two boost invariants [see Eq.~\eqref{eq:q2_z_delta}]. Due to the boost invariance in light-front dynamics, each of these frames represents infinitely many frames related by light-front boosts. Our work explores the full range of frame dependence that is implicit in the adopted light-front model for the systems investigated here. 

By analyzing the light-front wavefunction representation of the hadron matrix elements, we find that frames with minimal longitudinal momentum transfer could suppress non-valence contributions, and
are thus preferred for calculations in the valence Fock sector. Those are, the Drell-Yan frame in the space-like region, and the
longitudinal-II frame in the time-like region, as defined in Sec.~\ref{sec:TFF}. Our suggested frames agree with the study by Bakker and collaborators on the
semileptonic decay~\cite{Bakker:2003up}. In their work, the transition form factor in the time-like region obtained from one specific frame in
the valence Fock sector is closest to the full solution that also includes non-valence contributions. This specific frame is defined as
$\vec q_\perp=0$ with negative recoil, which resembles the longitudinal-II frame in this work.

As in our previous study on the radiative decays~\cite{Li:2018uif}, here we employ the light-front wave functions of the heavy quarkonia from the basis light-front quantization (BLFQ) approach~\cite{1stBLFQ}. The effective Hamiltonian is based on light-front holographic QCD and the one-gluon exchange light-front QCD interaction~\cite{Yang_fix,Yang_run}. 
We find that the frame dependence  of the transition form factor can be characterized as ranging between two limiting cases, similar to the dependence shown for the elastic form factor of (pseudo-)scalar mesons~\cite{Li:2017uug}. The spread of this frame dependence could serve as a measure for the violation of the Lorentz symmetry due to Fock space truncation.

The layout of this paper is as follows. In Sec.~\ref{sec:TFF}, we
introduce the formalism and methods to calculate the M1 transition
form factor with general frames on the light front. We then apply the
formalism to heavy quarkonia in the Basis Light Front Quantization (BLFQ) approach
in Sec.~\ref{sec:result} and there we present the results of the
transition form factors from different frames and the effective mass
spectrum for the resulting lepton pair in the Dalitz decay. We
summarize our paper in Sec.~\ref{sec:summary}.

\section{Transitions on the light front}
\label{sec:TFF}

\subsection{The decay width}

The Lorentz covariant decomposition of the electromagnetic transition matrix element between a vector meson ($\mathcal{V}$) and a pseudoscalar  ($\mathcal{P}$) is~\cite{Dudek_JPsi}

% \begin{align}
%   \bra{\mathcal{V}(P',m_j)} J^\mu(0) \ket{\mathcal{P}(P)}
%   =\frac{2 V(q^2)}{m_{\mathcal{P}}+m_{\mathcal{V}}}\epsilon^{\mu\alpha\beta\sigma} {P}_\alpha P'_\beta  e^*_{\sigma}(P', m_j)
%   \;,
% \end{align}

\begin{align}
  \bra{\mathcal{P}(P)} J^\mu(0) \ket{\mathcal{V}(P',m_j)}
  =\frac{2 V(q^2)}{m_{\mathcal{P}}+m_{\mathcal{V}}}\epsilon^{\mu\alpha\beta\sigma} {P}_\alpha P'_\beta e_{\sigma}(P', m_j)
  \;,
\end{align}
where $q^\mu = {P'}^\mu - P^\mu$ represents the momentum transfer between the initial and final mesons. On the light front, $\mu=+,-,x,y$ ($v^\pm=v^0\pm v^z$, we use the same conventions of the light-front coordinate as in Ref.~\cite{Li:2018uif}). $V(q^2)$ is the transition form factor. $m_{\mathcal{P}}$ and $m_{\mathcal{V}}$ are the masses of the pseudoscalar and the vector, respectively. $e_{\sigma}$ is the polarization vector of the vector meson, with $m_j(=0,\pm 1)$  the magnetic projection.
% $J^\mu(x)$ is the electromagnetic current operator.

%The matrix element for the radiative transition between a vector meson ($\mathcal{V}$) with four-momentum $P$ and polarization $m_j$ and a pseudoscalar  ($\mathcal{P}$) with four-momentum $P'$ via emission of a photon, can be parametrized in terms of the transition form factor $V(q^2)$ as~\cite{Dudek_JPsi},

In the process of $\psi_A\to\psi_B\gamma$, ($\psi_A, \psi_B = \mathcal V, \mathcal P$ or $\mathcal P, \mathcal V$), $q^2=0$, the decay width in the rest frame of the initial particle could be derived from the transition matrix element~\cite{Li:2018uif},
\begin{align}\label{eq:VPwidth}
  \begin{split}
    \Gamma(\psi_A\to\psi_B\gamma)
    =
    \frac{ {(m_A^2-m_B^2)}^3}{ {(2m_A)}^3{(m_A+m_B)}^2}
    \frac{{|V(0)|}^2}{(2 J_A+1)\pi}
    \;.
  \end{split}
\end{align}
 $J_A$ is the angular momentum of the initial meson $\psi_A$.
%({\color{red}Are there high order [ $O(\alpha_{em}^2)$] corrections to this expression?}{\color{blue}I think all orders of $\alpha_{em}$ is already included in this expression. $V(0)$ is defined from the transition amplitude of $\mathcal{V}\to\mathcal{P}$ in Eq. (1). We took the impulse approximation in later calculation, where $V(q^2)=2 e Q_f \hat{V}(q^2)$ and we calculate $\hat{V}(q^2)$. If higher order corrections are made to the current $J^\mu$, the transition form factor would be $V(q^2)=2 e Q_f \hat{V}(q^2) + \ldots$, where $\ldots$ stands for those corrections. })
% The amplitude of the Dalitz decay $\mathcal{V} \to \mathcal{P}+l^+ +l^-$ has the form
% \begin{align}\label{eq:D_amplitude}
%   \mathcal{M}_{m_j,\lambda}= \bra{\mathcal{V}(P',m_j)} J^\mu(0) \ket{\mathcal{P} (P)}\frac{1}{q^2} \bar u \gamma^\mu u,
% \end{align}
% where $\bar u \gamma^\mu u$ is the leptonic current.

For the Dalitz decay $\psi_A \to \psi_B l^+l^-$, the physical region of interest is $4 m_l^2 \le q^2\le (m_A- m_B)^2$.
The effective mass spectrum of the lepton pair could be derived as ~\cite{Landsberg:1986fd}:
\begin{align}\label{eq:D_width}
  \begin{split}
    \frac{\diff \Gamma(\psi_A \to \psi_B l^+ l^-)}{\diff q^2 \cdot \Gamma(\psi_A \to   \psi_B\gamma)}
    =\frac{\alpha}{3\pi} \sqrt{1-\frac{4 m_l^2}{q^2}}
\bigg( 1+ \frac{2 m_l^2}{q^2}\bigg)\frac{1}{q^2}&\\
\times \bigg[
\bigg(1+\frac{ q^2}{m_{A }^2-m_{B}^2}\bigg)^2
- \frac{ 4 m_{A }^2 q^2}{{(m_{A }^2-m_{B}^2)}^2}
\bigg]^{3/2}
    \bigg|\frac{V(q^2)}{V(0)}\bigg|^2&
    \;.
  \end{split}
\end{align}
% \begin{align}\label{eq:D_width}
%   \begin{split}
%     \frac{\diff \Gamma(\mathcal{V} \to   \mathcal{P}l^+ l^-)}{\diff q^2 \cdot \Gamma(\mathcal{V} \to   \mathcal{P}\gamma)}
%     =&\frac{\alpha}{3\pi} \sqrt{1-\frac{4 m_l^2}{q^2}}
% \bigg( 1+ \frac{2 m_l^2}{q^2}\bigg)\frac{1}{q^2}\\
% &\bigg[
% \bigg(1+\frac{ q^2}{m_{\mathcal{V} }^2-m_{\mathcal{P}}^2}\bigg)^2
% - \frac{ 4 m_{\mathcal{V} }^2 q^2}{{(m_{\mathcal{V} }^2-m_{\mathcal{P}}^2)}^2}
% \bigg]^{3/2}
%     \bigg|\frac{V(q^2)}{V(0)}\bigg|^2
%     \;.
%   \end{split}
% \end{align}

\subsection{Frames and kinematics}
\label{subsec:frame}

\begin{figure*}[htp]
  \subfigure[\ Regional plot of $q^2(\Delta_\perp, z)$ \label{fig:LFWF_rep_a}]
{\includegraphics[width=0.4\textwidth]{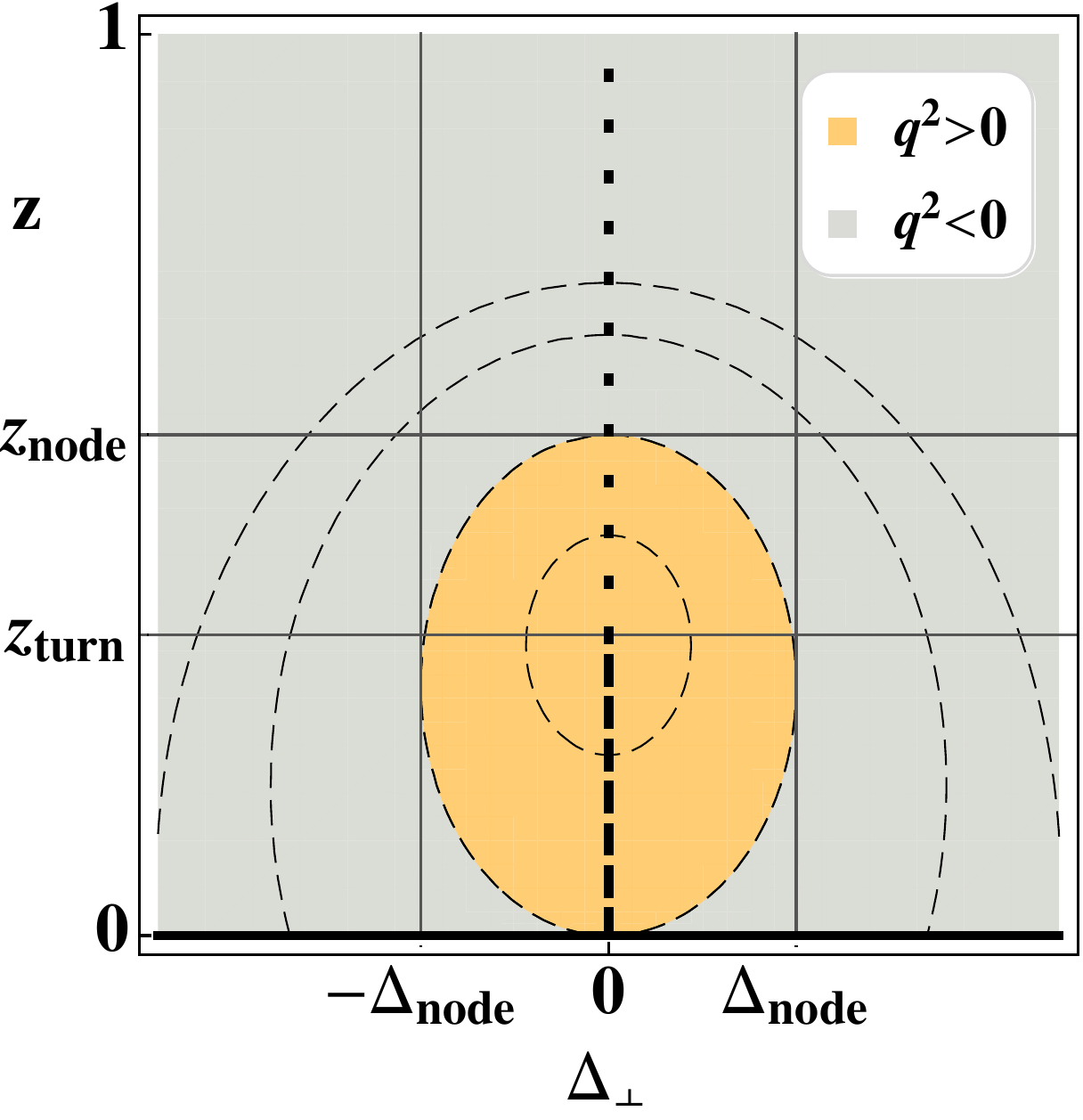}
} \qquad
  \subfigure[\ 3D plot of $q^2(\Delta_\perp, z)$ \label{fig:LFWF_rep_b}]
{\includegraphics[width=0.4\textwidth]{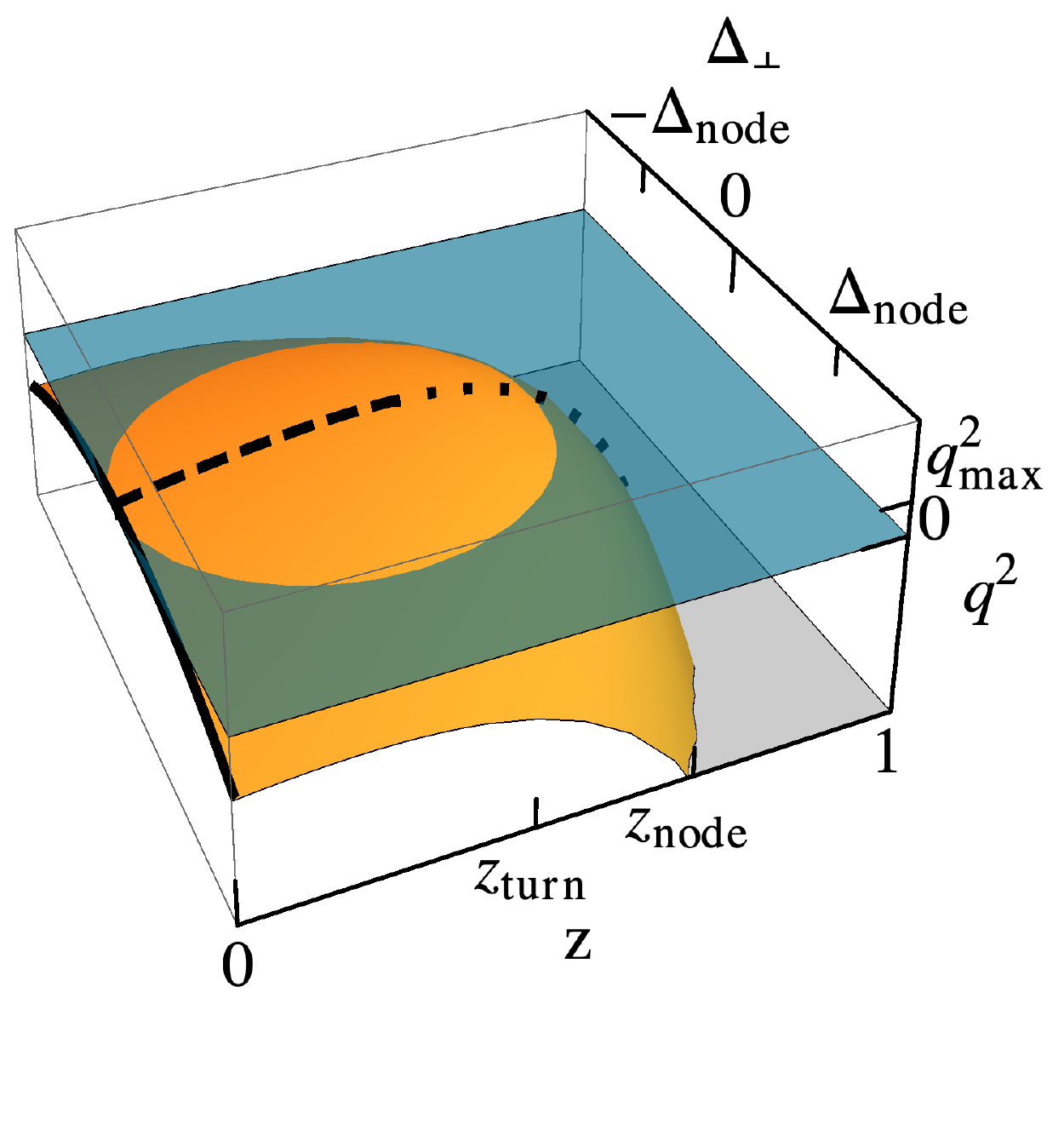}
}   
  \caption{\label{fig:q2_z_delta}Visualization of the Lorentz invariant momentum transfer squared $q^2$ as a function of $z$ and $\vec \Delta_\perp$ at $\arg \vec\Delta_\perp=0,\pi$. (a): regional plot of $q^2$. The time-like region ($q^2>0$) is the orange oval shape, bounded by $\Delta_{\text{node}}=(m^2_A-m^2_B)/2m_A$ and $z_{\text{node}}=1-m^2_B/m^2_A$. The space-like region ($q^2<0$) is in light gray. Contour lines of $q^2$ are indicated with thin dashed curves. The maximal value $q_{\max}^2=(m_A-m_B)^2$ occurs at $(z_{\text{turn}}=1-m_B/m_A, \Delta_\perp=0)$. (b): 3D plot of $q^2$ showing a convex shape in the $(z, \Delta_\perp)$ representation. The blue flat plane is the reference plane of $q^2=0$. In each figure, the Drell-Yan frame is shown as a thick solid line, and the longitudinal I and II frames are shown as thick dotted and thick dashed lines respectively.}
\end{figure*}

Considering the transition $\psi_A(P')\to\psi_B(P)\gamma^{(*)}(q=P'-P)$, the Lorentz invariant momentum transfer $q^2$ can be written as a function of two boost invariants~\cite{Li:2017uug} according to four-momentum conservation $q^2=(P'-P)^2$,
\begin{align}\label{eq:q2_z_delta}
q^2=zm_A^2-\frac{z}{1-z}m_B^2-\frac{1}{1-z}\vec\Delta_\perp^2\;.
\end{align}
where,
\[ z\equiv ({P'}^+-P^+)/{P'}^+, \qquad \vec \Delta_\perp\equiv\vec q_\perp-z\vec P'_\perp\;.\]
% \begin{figure}[htp]
% \includegraphics[width=0.4\textwidth]{kinematics_Yang}
%   \caption{\label{fig:kinematics_Yang}({\color{red}Yang: I made a diagram to show the kinematics. FYI.}
% {\color{blue} Thank you very much for making the figure. Could you explain on it? I am not quite sure what the red arc with $z,\vec\Delta_\perp$ mean.}
% {\color{red}Yang: red arc means that $z, \Delta_\perp$ are relative 
% momentum between the photon and the meson B. You don't have to include it in the paper.}
% {\color{cyan}Thanks for the information. I put it here for now so that Dr.Maris and Dr.Vary could also take a look. I will make some adjust to the labels if we decide include it in the paper.} )}
% \end{figure}
$z$ can be interpreted as the relative momentum transfer in the longitudinal direction, and $\vec \Delta_\perp$ describes the
momentum transfer in the transverse direction. Note that $z$ is restricted to $0\le z<1$ by definition.
%({\color{red}Yang: Is this really a restrictly? For example, if the photon is imcoming, z could be negative, right?}{\color{blue}This definition of $z$ is for the transition $\psi_A(P')\to\psi_B(P)\gamma^{(*)}(q=P'-P)$, as mentioned in the beginning of this section. For the incoming photon process, $\psi_A(P)\gamma^{(*)}(q=P'-P)\to\psi_B(P')$, if one take the same definition, $ z\equiv ({P'}^+-P^+)/{P'}^+$, it is still that $0\le z<1$. If you define $z$ as $ z\equiv ({P}^+-{P'}^+)/{P}^+$, then $z$ could be negative, but that is no longer the same $z$ as in the paper.}{\color{red}Yang: OK. That's not what I mean. I mean for the incoming process, you should consider $\psi_A(P')\gamma(q = P'-P) -> \psi_B(P)$. In this way, z could be negative. The point is that in covariant formulation, the incoming and outgoing processes are related by crossing symmetry. })
For each possible value of $q^2$, the values of the pair $(z,\vec\Delta_\perp)$ are not unique, and those different choices correspond to different reference
frames (up to longitudinal and transverse light-front boost transformations). Fig.~\ref{fig:q2_z_delta} should help visualize the functional form of $q^2(z,\vec\Delta_\perp)$. Since $q^2$ is relevant to the magnitude of $\vec\Delta_\perp$ but not its angle, we plot it in the $\arg \vec\Delta_\perp=0,\pi$ plane.
%The thin dashed contour lines in the left panel of Fig.~\ref{fig:LFWF_rep_a}. The right panel of Fig.~\ref{fig:LFWF_rep_b} presents a 3D plot of $q^2 (z,\vec\Delta_\perp)$ which is employed to visualize the various kinematic regions.
Transition form factors evaluated at different $(z, \vec\Delta_\perp)$ but at the same $q^2$ could reveal the frame dependence. In particular, we introduce two special frames for detailed consideration.
\begin{itemize}
\item Drell-Yan frame ($z=0$) :  $q^+=0$, $\vec \Delta_\perp = \vec q_\perp$  and $q^2=-\vec\Delta_\perp^2$. This frame is shown as a single thick solid line in each panel of Fig.~\ref{fig:q2_z_delta}. The Drell-Yan frame is conventionally used together with the plus
  current $J^+$ to calculate the electromagnetic form factors. This choice, on the one hand, avoids spurious effects related to the orientation of the null
  hyperplane where the light-front wavefunction is defined and, on the other hand, it suppresses the contributions from the often-neglected pair creation process, at least for pseudoscalar mesons \cite{CARBONELL1998215, DEMELO1998574, BRODSKY1999239, PhysRevD.65.094043, Simula:2002vm, PhysRevD.88.025036}. For the transition form factor, this is only true if zero-mode contributions are neglected. The
  transition form factor obtained in the Drell-Yan frame is significantly restricted in the space-like region, i.e. $q^2 \le 0$. Although one could analytically continuate the form factor to the time-like region by changing $\vec q_\perp$ to $i\vec q_\perp$
  ~\cite{Melikhov:1995xz, Jaus:1996np, Bakker:2003up}, we elect to calculate transition form factors directly from wavefunctions.
\item longitudinal frame ($\vec \Delta_\perp=0$): $q^2=zm_A^2-zm_B^2/(1-z)$. Note that we use the same definition for the
  longitudinal frame as in Ref.~\cite{Li:2017uug}, which is different from those in the literature where $\vec
  q_\perp=0$ is called the longitudinal frame~\cite{Isgur:1988iw, Sawicki:1992qj, BRODSKY1999239, Bakker:2003up}. In this frame, we have access to the
  kinematic region up to $q_{\max}^2=(m_A-m_B)^2$, the point where the final
  meson does not recoil. This maximal value occurs at $z = 1 - m_B/m_A \equiv z_{\text{turn}} $. For a given $q^2$, there are two solutions for $z$, corresponding to either positive or negative recoil direction of the final meson relative to the initial meson, namely,
%({\color{red}Usually, one thinks of the initial meson is at stationary. How does one define a reference direction from it?}{\color{blue}The direction is along the longitudinal direction. Since $\Delta_\perp=0$, the two different $z(= ({P'}^+-P^+)/{P'}^+)$ for the same $q^2$ correspond to the positive and negative recoil along the longitudinal.})
%({\color{red}Y.Li: I would suggest to make a 3D light cone plot to show the two kinematics. That will help the readers to understand the two choices and the frame dependence in general. }
%{\color{blue}M.Li: I made the 3D plot in Fig.2, do you mean something different from that?}
%{\color{red}Y.Li:They are not what I think of.}
%{\color{blue}M.Li: Could you explain more on the " 3D light cone plot''? I don't think I get it. } )
% \begin{figure}[htp]
%  \includegraphics[width=0.3\textwidth]{frame_LC_Yang.png}  
%   \caption{\label{fig:frame_LC}Visualization of frames.}
% \end{figure}
\begin{itemize}
\item longitudinal-I: $z=[m_A^2-m_B^2+q^2+\sqrt{(m_A^2-m_B^2+q^2)^2-4m_A^2q^2}]/(2m_A^2)$. $ z_{\text{turn}}\le z < 1$. This branch joins the second branch at
  $q^2=q^2_{\max}$ with $z=z_{\text{turn}},\vec\Delta_\perp=0$. The time-like region is accessed at $z_{\text{turn}}\le z <
  z_{\text{node}}$, and the space-like region is at $ z_{\text{node}}\le z < 1$, where $z_{\text{node}}\equiv 1-m^2_B/m^2_A$. The longitudinal-I frame is shown as thick dotted lines in Fig.~\ref{fig:q2_z_delta}.
\item longitudinal-II:
$z=[m_A^2-m_B^2+q^2-\sqrt{(m_A^2-m_B^2+q^2)^2-4m_A^2q^2}]/(2m_A^2)$.
$0\le z\le z_{\text{turn}}$. This second branch only exists in the time-like region, and it joins the Drell-Yan frame at $q^2=0$
with $z=0,\vec\Delta_\perp=0$. The longitudinal-II frame is shown as thick dashed lines in Fig.~\ref{fig:q2_z_delta}.
\end{itemize}
\end{itemize}

\subsection{Light-front wavefunction representation of the M1 transition form factor}

The meson state vector $ \ket{\psi_h(P,j,m_j)}$ can be expanded in the light-front Fock space. The coefficients of the Fock expansion are the complete set
of n-particle light-front wavefunctions (LFWFs),  $\{\psi_{n/h}^{(m_j)}(x_i,\vec k_{i\perp}, s_i)\}$. $x_i\equiv
{\kappa_i^+}/{P^+}$ is the longitudinal momentum fraction of the i-th parton, and $\vec k_{i\perp}\equiv \vec \kappa_{i\perp}-x \vec
P_\perp$ is the relative transverse momenta, with $\kappa_i$ being the momenta of the corresponding parton. $s$ is the spin of the parton.

% \begin{align}
%   \begin{split}
%     % relative coordinates
%  \ket{\psi_h(P,j,m_j)}&
%    =\sum_n    f_n \sum_{s_i,l_i}
%    \prod_{i=1}^n\int\frac{\diff x_i \diff^2 k_{i\perp}}{{(2\pi)}^32 x_i}\\
%    &\times   2{(2\pi)}^3
%     \delta(\sum_{i=1}^n x_i- 1)
% \delta^{(2)}(\sum_{i=1}^n\vec k_{i\perp})\\
% &\times \psi_{n/h}^{(m_j)}(\{x_i,\vec k_{i\perp}, s_i\})
%     c^\dagger_{s_1 l_1}(\kappa_1)\cdots c^\dagger_{s_n l_n}(\kappa_n)\ket{0}
%     \; ,
%   \end{split}
% \end{align}
% where  $c^\dagger_{s l}(\kappa)$ is the creation operator for the corresponding constituent (quark, anti-quark or gluon) with spin
% $s$ and color $l$, and $f_n$ is the color factor for the n-parton sector.

\begin{figure*}[ht!]
   \subfigure[\  $\psi^{(m_j=0)}_{\uparrow\downarrow+\downarrow\uparrow/J/\psi}(x'= x+z(1-x), \vec k'_\perp=\vec k_\perp +(1-x)\vec\Delta_\perp)$ \label{fig:LFWF_jpsi_shifts}]
   {\includegraphics[width=0.22\textwidth]{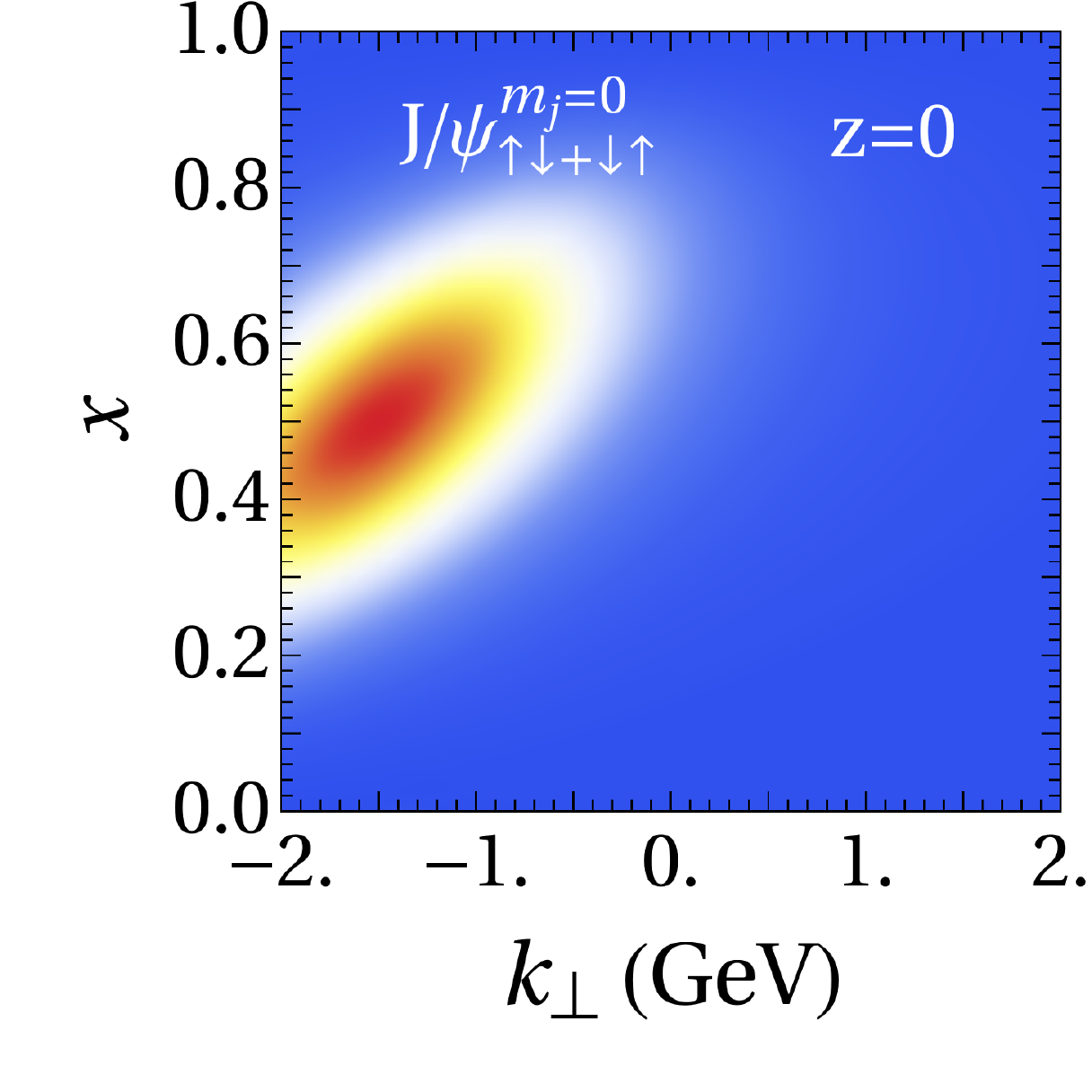}\hspace{-.3cm}
     \includegraphics[width=0.22\textwidth]{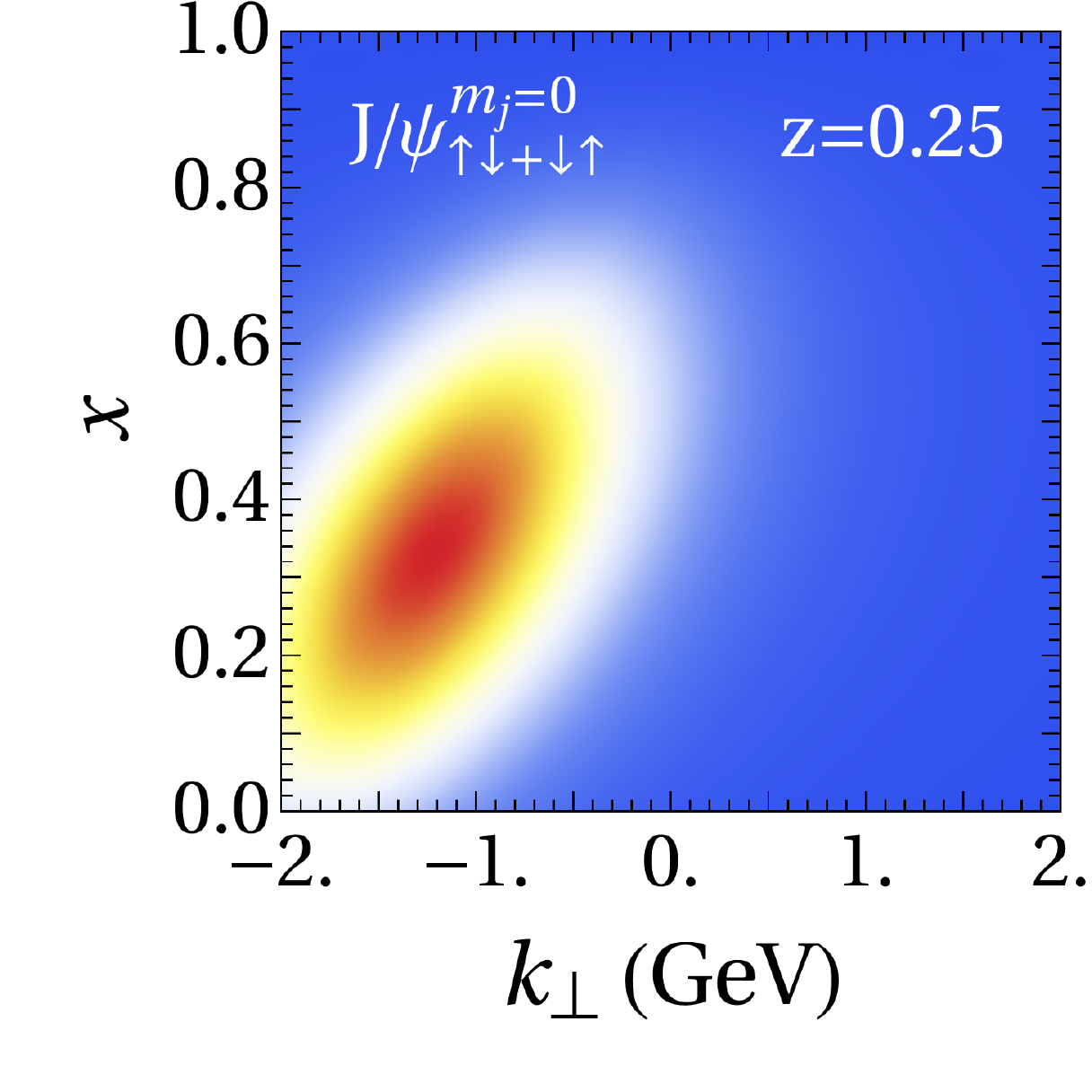}\hspace{-.3cm}
     \includegraphics[width=0.22\textwidth]{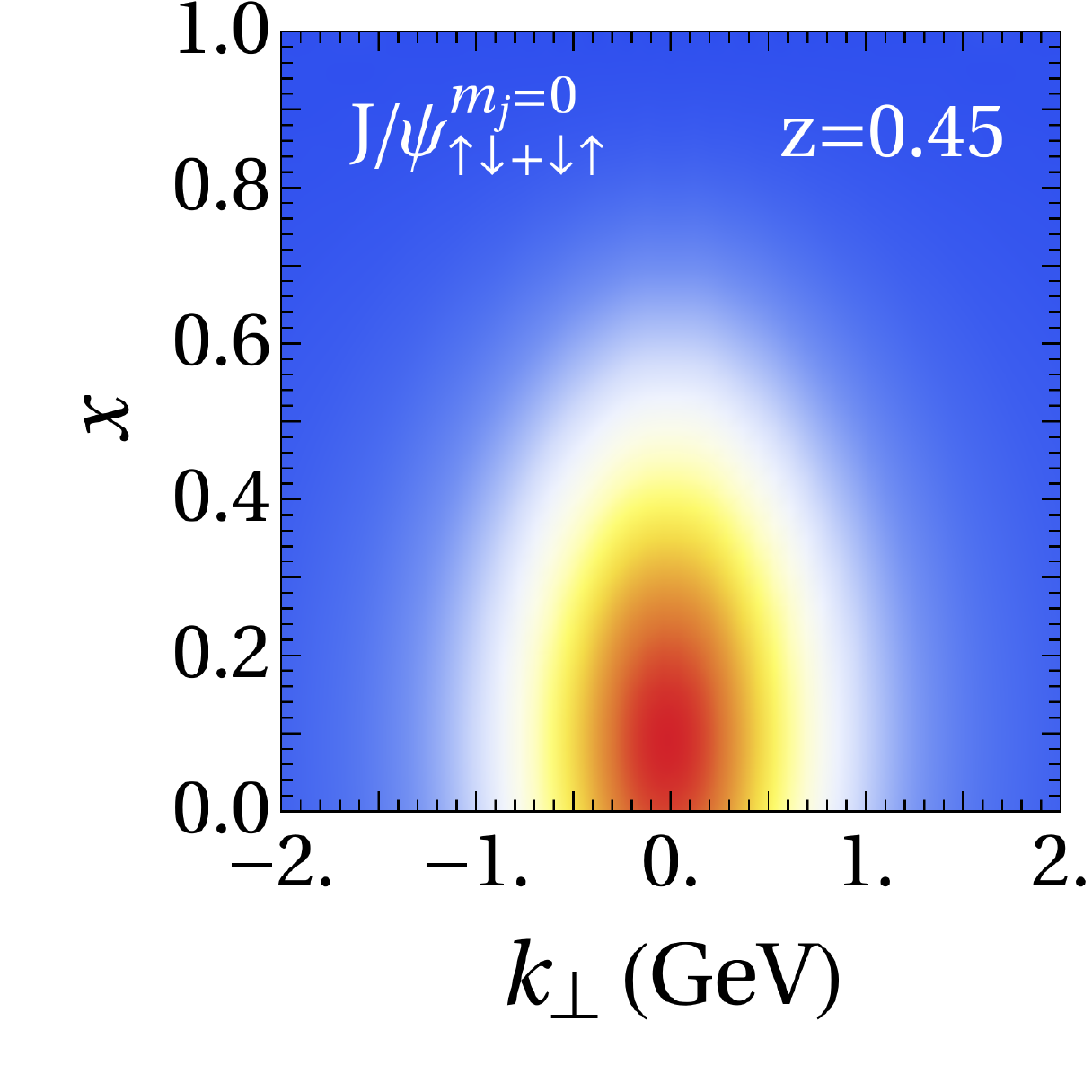}
   } \qquad
   \subfigure[\ $\psi^{(m_j=0)}_{\uparrow\downarrow-\downarrow\uparrow/\eta_c}(x, \vec k_\perp)$ \label{fig:LFWF_etac}]
   {\includegraphics[width=0.22\textwidth]{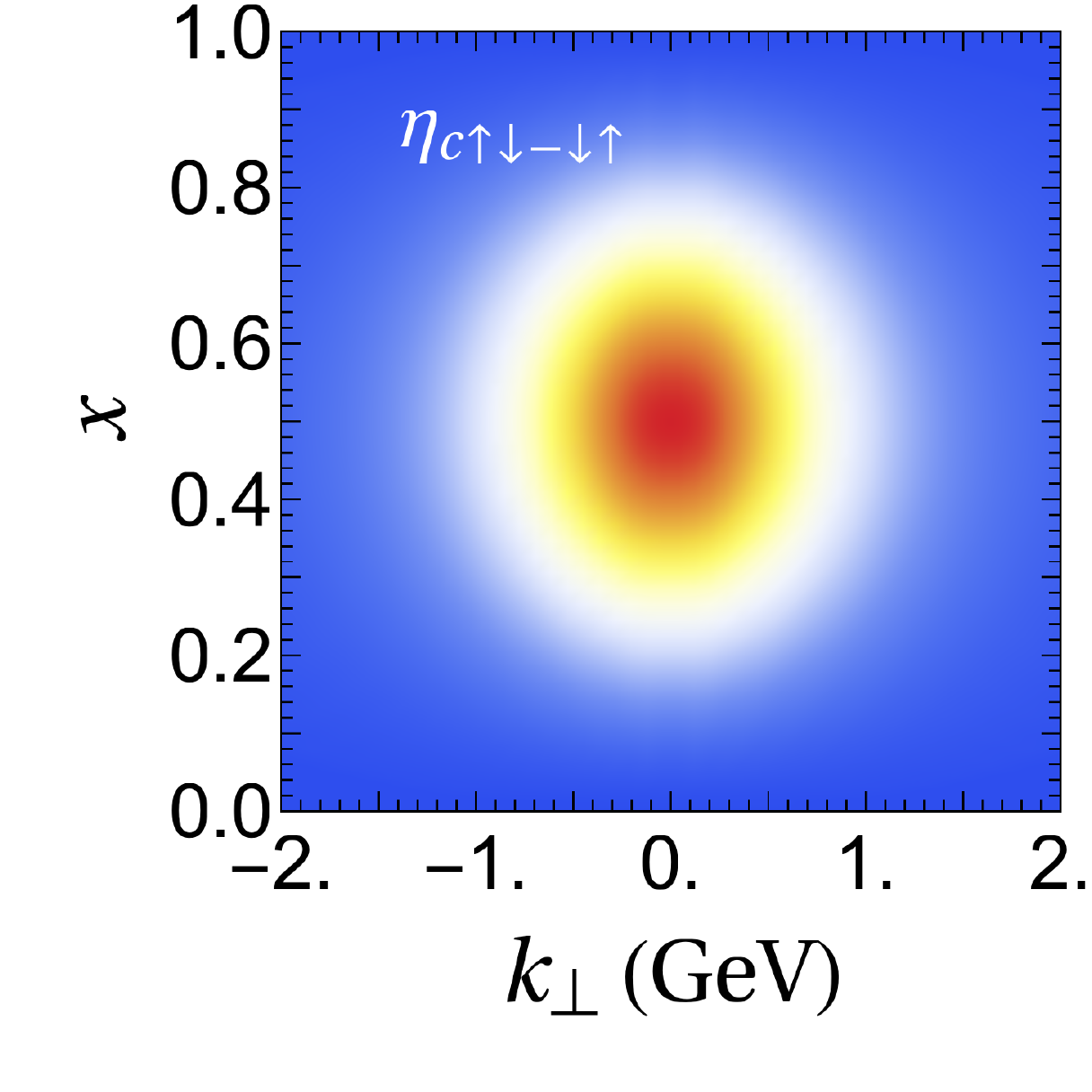}
     \includegraphics[width=0.035\textwidth]{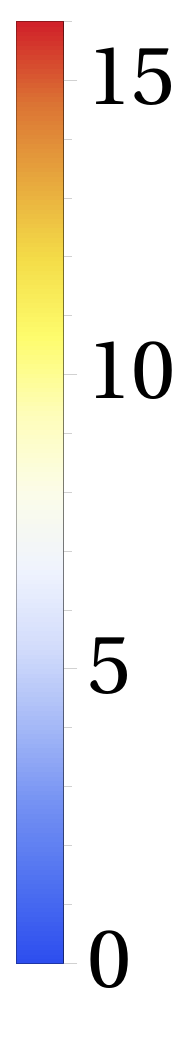}
   }   
   \caption{\label{fig:LFWF_overlap}The valence light-front wavefunctions of mesons as they contribute (see Eq.~\eqref{eq:HadronMatrix_nn}) to the convolution in the transition $J/\psi\to
     \eta_c+\gamma^{(*)}$ at $q^2=-3 \GeV^2$ in different frames. According to Eq.~\eqref{eq:HadronMatrix_nn}, in this $2\to 2$
     parton-number-conserving term, the initial state
     wavefunction of $J/\psi$ would appear shifted and stretched to overlap with the final state wavefunction of $\eta_c$, when
     plotted on the $(x,\vec k_\perp)$ space. Shown in (a), the wavefunction
     of  $J/\psi$ is shaped differently at different  $(z,\vec \Delta_\perp)$, i.e. in different frames. The longitudinal
     dimension is preserved most in the Drell-Yan frame where $z=0$. At larger $z$, the information in the longitudinal region is
     reduced, and the transverse shift becomes smaller. The largest $z$ is achieved when $\Delta_\perp=0$ in the longitudinal frame,
     in this case, $z=0.45$. Plotted in (b) is the wavefunction of $\eta_c$. All light-front wavefunctions what we employ are calculated by
     Ref.~\cite{Yang_run} and here we only plot the dominant spin components for the purpose of illustration.}
 \end{figure*}

The transition amplitude $\psi_A\to\psi_B\gamma^{(*)}$ is given by the sum of the diagonal $n\to n$ and off-diagonal $n+2\to
n$ transitions, as shown in Fig.~\ref{fig:partonN}. 
For the $n\to n$ term, as in Fig.~\ref{fig:nton}, the external photon is coupled to a quark or an antiquark, thus the
electromagnetic current matrix element takes the form
\begin{widetext}
\begin{align}\label{eq:HadronMatrix_nn}
    \begin{split}
      \bra{\psi_{B}(P,j,m_j)} J^\mu(0) \ket{\psi_{A}(P',j',m_j')}_{n\to n}
      =&\sum_n f_n^2\prod^n_{i=1}\sum_{s_i',s_1,l_i'}
\int_z^1\frac{\diff x'_1 }{2 x'_1}
      \int_0^1 \frac{\diff x'_{i(i\ne 1)} }{2 x'_i}
      \int\frac{\diff^2 k'_{i\perp}}{{(2\pi)}^3}
        2{(2\pi)}^3
      \delta(\sum_{i=1}^n x_i'- 1)
      \delta^{(2)}(\sum_{i=1}^n\vec k_{i\perp}')\\
      &\times
      \psi_{n/B}^{(m_j)*}(\{x_i,\vec k_{i\perp}, s_i\})
      j_{s_1,s_1'}^\mu
      \psi_{n/A}^{(m'_j)}(\{x'_i,\vec k'_{i\perp}, s'_i\})
      \;,
    \end{split}
  \end{align}
where the EM current 
  \(
  j_{s_1,s_1'}^\mu=
  \bar u_{s_1'}(\kappa_1')
  \gamma^\mu
  u_{s_1}(\kappa_1)
  \) if the struck parton is a quark, and \(
  j_{s_1,s_1'}^\mu=
  \bar v_{s_1}(\kappa_1)
  \gamma^\mu
  v_{s_1'}(\kappa_1')
  \) if the struck parton is an antiquark. $l_i$ is the color index of the i-th parton and $f_n$ is the color factor for the n-parton sector.
  The relative coordinates and constraint conditions of partons are 
\begin{align}
  \begin{cases}
     x_1'= x_1 + z(1- x_1), \vec k'_{1\perp}=\vec k_{1\perp}+(1-x_1) \vec \Delta_\perp,  l_1=l'_1,
    & \text{for the struck parton } (i=1)\\
 x_i'= x_i (1-z), \vec k'_{i\perp}=\vec k_{i\perp}-x_i\vec \Delta_\perp,  l_i=l_i',  s_i=s_i',
& \text{for the spectators } (i=2,\ldots, n)\;.
  \end{cases}
\end{align}
\end{widetext}
 In Sec.~\ref{subsec:frame}, we have shown that different choices of $(z,\vec \Delta_\perp)$
 for the same $q^2$ could characterize different frames. Consider the $z$ which is the lower bound of the range of $x_1'$ for the $n\to n$ matrix element
 in Eq.~\ref{eq:HadronMatrix_nn}. As a consequence, increasing the value of $z$ would reduce the overlap region of the two wavefunctions in the
 longitudinal direction. We illustrate this effect in Fig.~\ref{fig:LFWF_overlap} by visualizing the convoluted valence wavefunctions~\cite{Yang_run} at different  $(z,\vec
 \Delta_\perp)$ with the same $q^2$ for the transition $J/\psi\to \eta_c+\gamma^{(*)}$. In the valence Fock sector,  the
 light-front wavefunction can be written in the form of $\psi^{(m_j)}_{s\bar s/h}(\vec k_\perp, x)$ where $(x,\vec k_\perp)$ is the
 relative coordinates of the quark. $s$ represents the fermion spin projection in the $x^-$ direction. 
 Both the initial and final wavefunctions
 are plotted in the same $(x,\vec k_\perp)$ space, where the initial state
 wavefunction would appear as being reshaped due to momentum transfer. We can see that the information from the wavefunction in the longitudinal direction is preserved most at minimal $z$.
 
For the $n+2\to n$ term, as in Fig.~\ref{fig:np2ton}, a quark and an antiquark from the initial state annihilate into a photon, thus the
electromagnetic current matrix element takes the form
\begin{widetext}
\begin{align}
  \begin{split}
    \bra{\psi_{B}(P,j,m_j)} J^\mu\ket{\psi_{A}(P',j',m_j')}_{n+2\to n}
    =&\sum_n f_n f_{n+2}\prod^{n+2}_{i=1}\sum_{s_i',l_i'}
\int_0^z\frac{\diff x'_1 }{2 x'_1}
    \int_0^1 \frac{\diff x'_{i(i\ne 1,2)} }{2 x'_i}
    \int\frac{\diff^2 k'_{i\perp}}{{(2\pi)}^3}
2{(2\pi)}^3
    \delta(\sum_{i=1}^{n+2} x_i'- 1)
    \delta^{(2)}(\sum_{i=1}^{n+2}\vec k_{i\perp}')\\
    &\times
    \psi_{n/B}^{(m_j)*}(\{x_i,\vec k_{i\perp}, s_i\})
    j_{s_1',s_2'}^\mu
    \psi_{n+2/A}^{(m'_j)}(\{x'_i,\vec k'_{i\perp}, s'_i\})
    \;,
  \end{split}
\end{align}
where the EM current is
\(j_{s_1',s_2'}^\mu = \bar v_{s'_1}(\kappa_2')
\gamma^\mu
u_{s'_2}(\kappa_1')
\)
and the parton coordinates/constraints are
\begin{align}
  \begin{cases}
    x'_1,\vec k'_1, &\text{for the struck quark}\\
    x'_2 =z- x'_1 , \vec k'_2=-\vec k'_1+\vec \Delta_\perp,  l_2'=l_1' ,&
    \text{for the struck anti-quark}\\
    x'_{i+2} =x_i(1-z), \vec k'_{i+2\perp}=\vec k_{i \perp}-x_{i}\vec \Delta_\perp,  l_i=l'_{i+2},  s_i=s'_{i+2}, &
    \text{for the spectators} (i=1,\ldots, n)\;.
  \end{cases}
\end{align}
\end{widetext}
The frame parameter $z$ is now the upper bound of the range of $x_1'$, suggesting that decreasing the value of $z$ might reduce the
contribution of the $n+2\to n$ transition
to the full transition form factor. However, even at $z=0$, this parton-number-non-conserving term may yield a non-zero value, by generating zero-mode $\delta(x)$ terms~\cite{Chang:1972xt,Burkardt:1989wy,DEMELO1998574,BRODSKY1999239}. In the space-like region, the Drell-Yan frame always has the minimal $z=0$. In
the time-like region, it is the longitudinal-II frame that takes the smallest $z$. When the $n+2\to n$ contribution is not
accessible, which happens when the light-front wavefunctions are solved in a truncated Fock space, using those minimal-$z$ frames
seems advantageous in suppressing the parton-number-non-conserving contribution. This observation suggests optimal frames for our meson systems solved from light-front
Hamiltonian in the valence Fock sector.

In the valence Fock sector, the transition amplitude of $\psi_A\to \psi_B$ with the current operator $J^\mu(x) = \overline \psi(x)
\gamma^\mu \psi(x)$ now contains only two contributions, one from the photon coupling to the quark and the other from the photon
coupling to the antiquark. For quarkonia,  the two terms are equal by charge conjugation. For convenience, in calculating the transition form factor, we only consider the photon coupling to the quark and the charge is not included. That is, we compute $\hat{V}(q^2)$ which is related to $V(q^2)$ by $\hat{V}(q^2) = V(q^2)/(2e \mathcal{Q}_f)$ where $\mathcal{Q}_f $ is
the dimensionless fractional charge of the quark.

In our previous work~\cite{Li:2018uif}, we have shown that for calculations with light-front wavefunctions in the valence Fock sector, using the transverse current $J^R(\equiv J^x + iJ^y)$ with the $m_j=0$ state of the vector meson is
preferred for the transition form factor $V(q^2)$, as in Eq.~\eqref{eq:VR_mj0}. We will adopt this choice for the purpose of studying the frame dependence in this work. Therefore, we employ
% \begin{align}\label{eq:VR_mj0}
%   \bra{\mathcal{V}(P',m_j)} J^R(0) \ket{\mathcal{P} (P)}
%   =\frac{2 V(q^2)}{m_{\mathcal{P}}+m_{\mathcal{V}}}im_{\mathcal{V}}{P}^+\bigg[ \dfrac{{P'}^R}{{P'}^+}-\dfrac{P^R}{P^+} \bigg]
%   \;.
% \end{align}
\begin{align}\label{eq:VR_mj0}
  \bra{\mathcal{P} (P)} J^R(0) \ket{\mathcal{V}(P',m_j)}
  =\frac{2 V(q^2)}{m_{\mathcal{P}}+m_{\mathcal{V}}}im_{\mathcal{V}}{P}^+\bigg[ \dfrac{P^R}{P^+}-\dfrac{{P'}^R}{{P'}^+} \bigg]
  \;.
\end{align}

The light-front wavefunction representation of the transition form factor $\hat V(q^2)$, extracted according to the expression in Eq.~\eqref{eq:VR_mj0}, follows as
%   \begin{align}\label{eq:Vmj0_red}
%     \begin{split}
%       \hat V(q^2)
%       =
%       &-i \frac{m_{\mathcal{V}}+m_{\mathcal{P}}}{2m_{\mathcal{V}}\Delta^R}\sum_{\bar s}
%       \int_0^1\frac{\diff x}{2x(1-x)}
%       \int\frac{\diff^2\vec k_\perp}{{(2\pi)}^3}
%       \frac{2}{\sqrt{x(1-z)[x+z(1-x)]^3}}
%       \\
%       &\times \Big[
%       \psi_{\uparrow \bar s/\mathcal{P}}(\vec k_\perp, x)
%       \psi_{\uparrow \bar s/\mathcal{V}}^{(m_j=0) *}(\vec k'_\perp, x')
%       (zk^R -x\Delta^R)
%       +\psi_{\downarrow  \bar s/\mathcal{P}}(\vec k_\perp, x)
%       \psi_{\uparrow \bar s/\mathcal{V}}^{(m_j=0) *}(\vec k'_\perp, x')
% m_qz
%       \Big]\;,
%     \end{split}
%   \end{align}
  \begin{align}\label{eq:Vmj0_red}
    \begin{split}
      \hat V(q^2)
      =
      &-i \frac{m_{\mathcal{V}}+m_{\mathcal{P}}}{2m_{\mathcal{V}}\Delta^R}\sum_{\bar s}
      \int_0^1\frac{\diff x}{2x(1-x)}
      \int\frac{\diff^2\vec k_\perp}{{(2\pi)}^3}\\
      &\frac{2}{\sqrt{x(1-z)[x+z(1-x)]^3}}
      \Big[
      \psi^*_{\uparrow \bar s/\mathcal{P}}(\vec k_\perp, x)
      \psi_{\uparrow \bar s/\mathcal{V}}^{(m_j=0) }(\vec k'_\perp, x')\\
      \\
      &\times (zk^R -x\Delta^R)
      +
     \psi^*_{\downarrow  \bar s/\mathcal{P}}(\vec k_\perp, x)
     \psi_{\uparrow \bar s/\mathcal{V}}^{(m_j=0) }(\vec k'_\perp, x')
     m_qz
      \Big]\;.
    \end{split}
  \end{align}
%where we have used the symmetries in light-front wavefunctions,
%$\psi_{\downarrow\uparrow/\mathcal{V}}^{(m_j=0)}=\psi_{\uparrow\downarrow/\mathcal{V}}^{(m_j=0)}$ and
%$\psi_{\downarrow\uparrow/\mathcal{P}}=-\psi_{\uparrow\downarrow/\mathcal{P}}$.
%We can see from Eq.~\eqref{eq:Vmj0_red} that, for different frames, i.e.\ different $\{z, \vec\Delta_\perp\}$, the convolution of light-front wavefunctions are evaluated with different shifts and scales.
We will numerically probe the frame dependence of the transition form factor through a dense sampling of kinematically available frames within $(z, \vec\Delta_\perp)$
for any given $q^2$. We will also try to see if our suggested frames, those with minimal-$z$, provide a better results than other frames,
when compared with available experimental data.

%In principle, a precise evaluation of the transition form factor requires knowledge of the full light-front wavefunctions and should not depend on the frame.
%Note that the $\Delta_R$ in the denominator does not cause singularity by exact cancellation with that in the wavefunctions at the numerator.
%In the Drell-Yan frame($z=0$), the terms that are proportional to $z$ would vanish. In the longitudinal frame($\Delta_\perp=0$),
 % \begin{align}\label{eq:Vmj0_DY}
 %    \begin{split}
 %      \hat V_{DY}(q^2)
 %      =
 %      &i \frac{m_{\mathcal{V}}+m_{\mathcal{P}}}{m_{\mathcal{V}}}\sum_{\bar s}
 %      \int_0^1\frac{\diff x}{2x(1-x)}
 %      \int\frac{\diff^2\vec k_\perp}{{(2\pi)}^3}
 %      \frac{1}{x}
 %      \psi_{\uparrow \bar s/\mathcal{P}}(\vec k_\perp, x)
 %      \psi_{\uparrow \bar s/\mathcal{V}}^{(m_j=0) *}(\vec k'_\perp, x')
 %     \;.
 %    \end{split}
 %  \end{align}

\section{ Results: the M1 transitions in heavy quarkonia}\label{sec:result}
We adopt light-front wavefunctions of heavy quarkonia from recent works~\cite{Yang_fix, Yang_run} in the BLFQ approach~\cite{1stBLFQ}. The
effective Hamiltonian extends the holographic QCD~\cite{holography} by introducing the one-gluon exchange interaction with a
running coupling, the constituent masses for the quarks and a longitudinal confining interaction~\cite{positronium}. The obtained light-front
wavefunctions have been used to produce several observables and are in reasonable agreement with experiments and with other theoretical
approaches~\cite{Chen:VM, sofia, Li:2017uug,Lekha_EFF, Chen:2018vdw}. We have also used these light-front wavefunctions to address radiative transitions in a previous work~\cite{Li:2018uif}. The previous work studied the transition form factor in the space-like region in the Drell-Yan frame, and suggested a preferred current component for practical calculations. The present work extends the calculation to the full kinematic region and all possible frame choices. 

In this model, the light-front wavefunctions are solved in the valence Fock sector using a basis function
representation, where the truncations in the Fock space and the basis space could introduce a violation of Lorentz symmetry. However, such a violation
turned out to be very small in terms of the meson mass spread at different magnetic projections, as well as in the elastic form factors for
mesons at different frames~\cite{Li:2017uug}. It is therefore worthwhile to 
examine how the transition form factors exhibit Lorentz symmetry violation as measured by their frame-dependence.

The light-front wavefunction in a basis function representation reads:
\begin{equation}\label{eq:LFWF}
  \psi^{(m_j)}_{s\bar s/h}(\vec k_\perp, x) = \sum_{n, m, l} \psi_h(n, m, l, s, \bar s) \, \phi_{nm}(\frac{\vec k_\perp}{\sqrt{x(1-x)}}) \chi_l(x).
\end{equation}
In the transverse direction, the 2D harmonic oscillator (HO) function $\phi_{nm}$ is adopted as the basis. In the longitudinal
direction, we use the modified Jacobi polynomial $\chi_l$ as the basis. $m$ is the orbital angular momentum projection, related to the
total angular momentum projection as $m_j=m+s+\bar s$, which is conserved by the Hamiltonian. The basis space is truncated by
their reference energies in dimensionless units:
\(2n+|m|+1\le N_{\max}\) and \( 0\le l \le L_{\max}\). As such, the $N_{\max}$-truncation provides a natural pair of UV and IR cutoffs: $\Lambda_{\perp,\textsc{uv}} \simeq \kappa\sqrt{N_{\max}}$,
$\lambda_{\perp,\textsc{ir}} \simeq \kappa/\sqrt{N_{\max}}$, where $\kappa$ is the oscillator basis energy scale parameter as well as the confining strength parameter. $L_{\max}$ represents the
resolution of the basis in the longitudinal direction $\Delta x \approx L_{\max}^{-1}$. It also provides a pair of UV and IR cutoffs according to Eq.~\eqref{eq:q2_z_delta}: $\Lambda_{z,\textsc{uv}} \simeq m_h\sqrt{L_{\max}}$,
$\lambda_{z,\textsc{ir}} \simeq m_h/\sqrt{L_{\max}}$ ($m_h\approx m_A,m_B$).
 See Ref.~\cite{Yang_run} for details on basis functions and parameter values. The light-front wavefunctions are calculated at $N_{\max}=L_{\max}=8,16,24$ and $32$. For our purposes, we mainly concentrate on results obtained at $N_{\max}=L_{\max}=32$. In this basis, the largest supported $|q^2|$ is 
$31~\GeV^2 (44~\GeV^2)$ for charmonia (bottomonia), and beyond these cutoffs, the LFWFs are dominated by the asymptotics of the basis. 
\subsection{The transition form factors in different frames}
\begin{figure*}[ht]
  \centering
  \includegraphics[height=6cm]{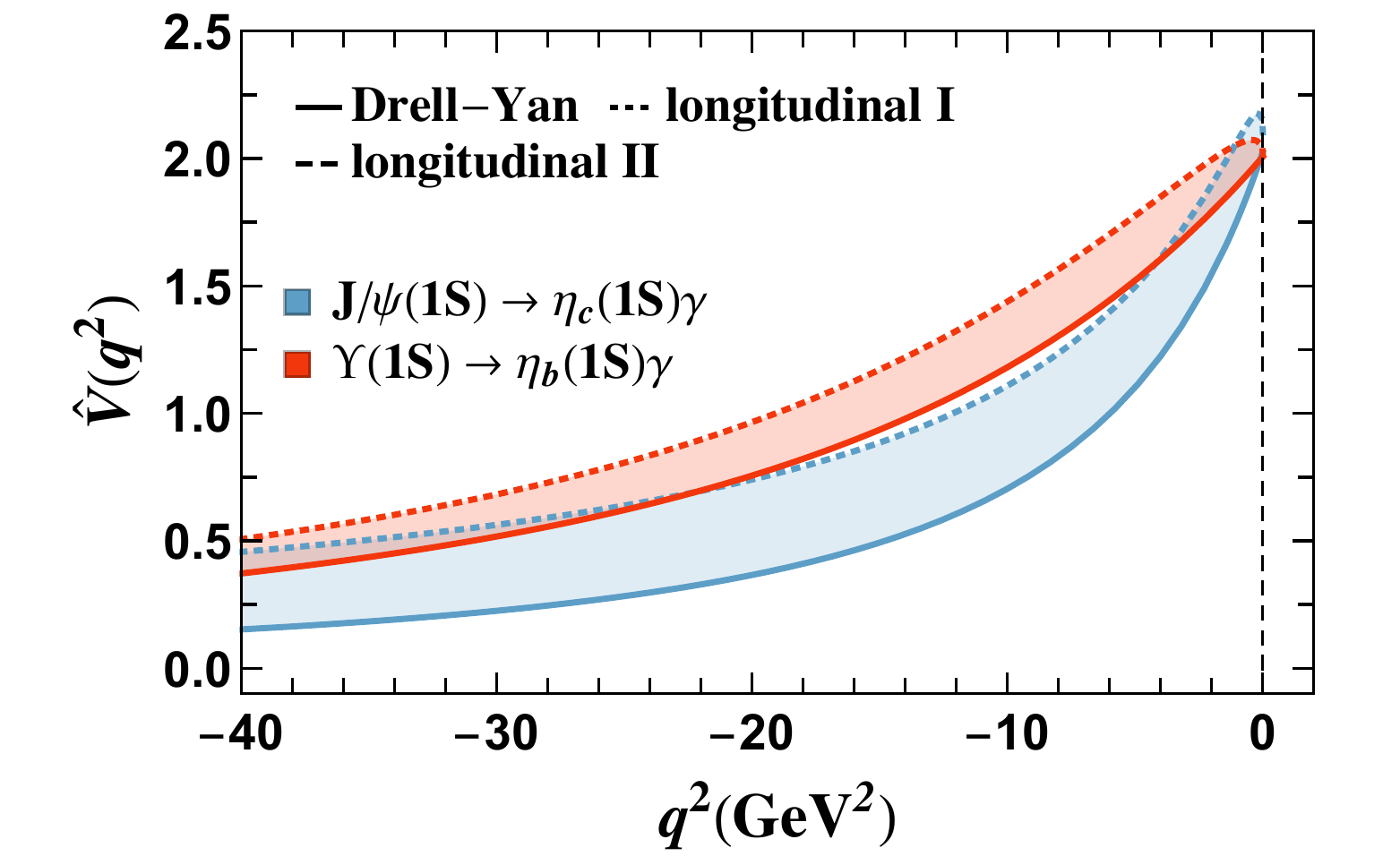}
  \includegraphics[height=6cm]{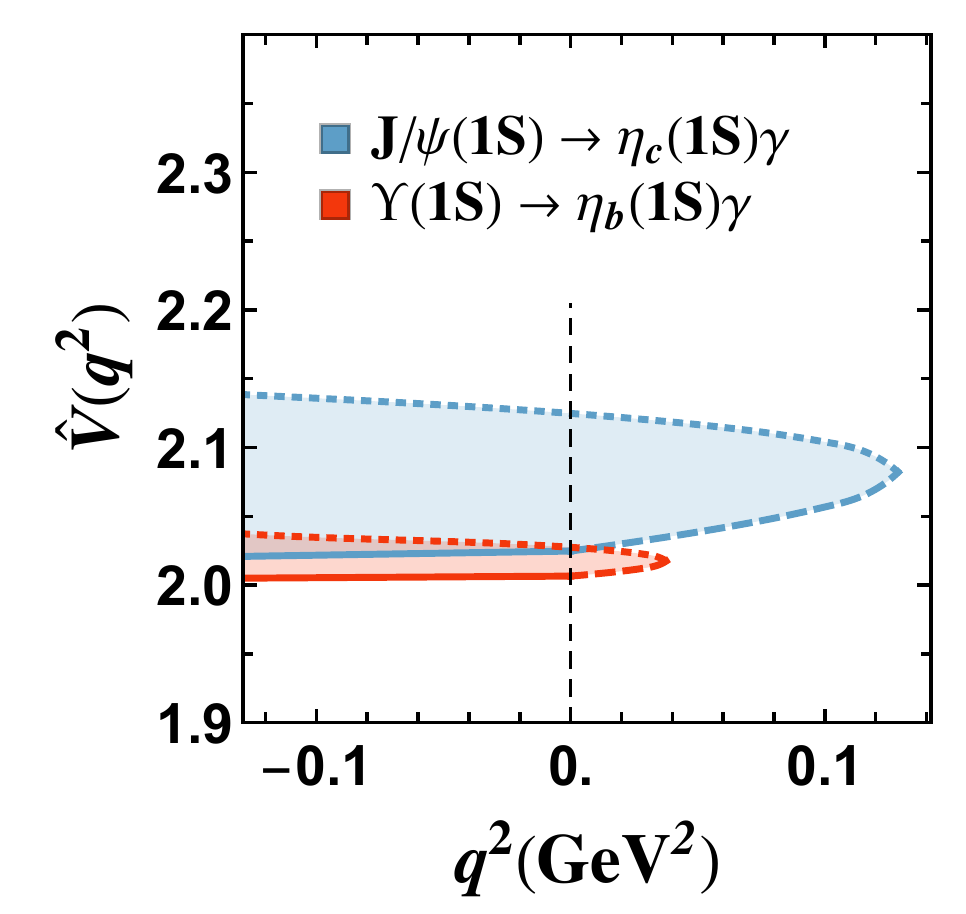}

  \includegraphics[height=6cm]{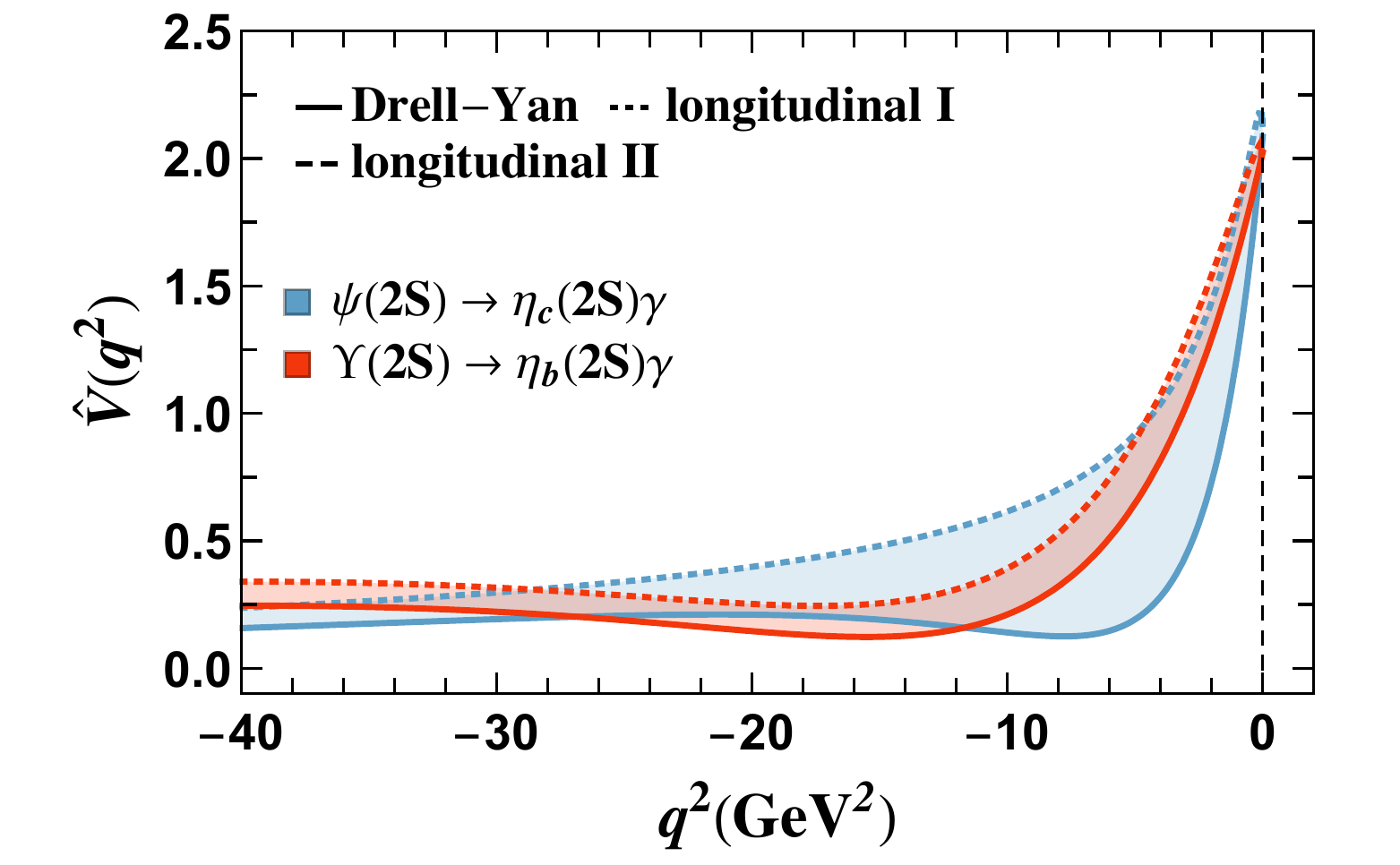}
  \includegraphics[height=6cm]{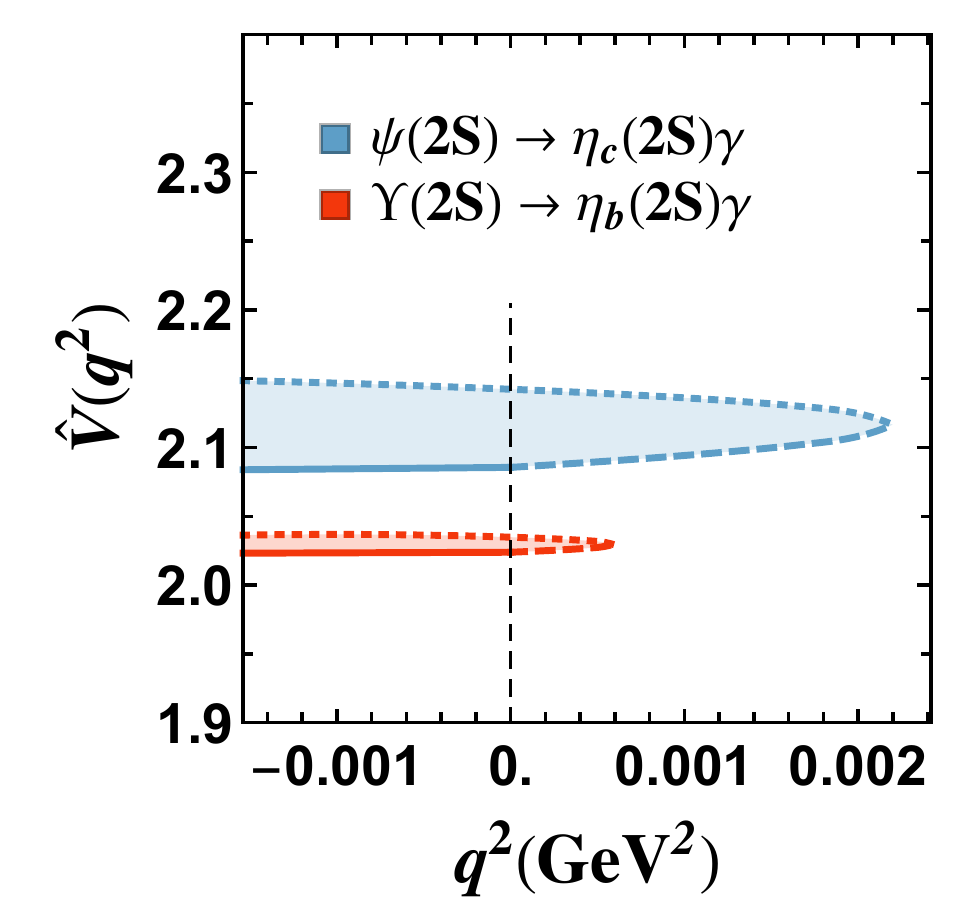}

  \caption{The transition form factor of the transition $\mathcal{V}(nS)\to\mathcal{P}( nS)\gamma$ of charmonia (blue curves/shades) and bottomonia (red curves/shades), calculated with light-front wavefunctions at $N_{\max} = L_{\max} = 32$ basis truncation. Meson masses are taken from experimental data~\cite{PDG2018} in defining the frames according to Eq.~\eqref{eq:q2_z_delta}. The solid curves represent the Drell-Yan frame while the other curves represent the longitudinal I (dotted lines) and II (dashed lines) frames. The shaded areas represent the results from all other frames. The left panel shows the transition form factor at a larger scale of $q^2$, and the right panel focuses on the small $q^2$ region. 
%Values of $\hat V(0)$ converted from available decay widths in PDG~\cite{PDG2018}: $\hat V(0)|_{J/\psi(1S)\to\eta_c(1S)\gamma}= 1.56(19)$, $\hat V(0)|_{\Psi(2S)\to\eta_c(2S)\gamma}=2.52(91)$.
}
 \label{fig:TFFccbb1}
\end{figure*}

\begin{figure*}[htp!]
  \centering
  \includegraphics[height=6cm]{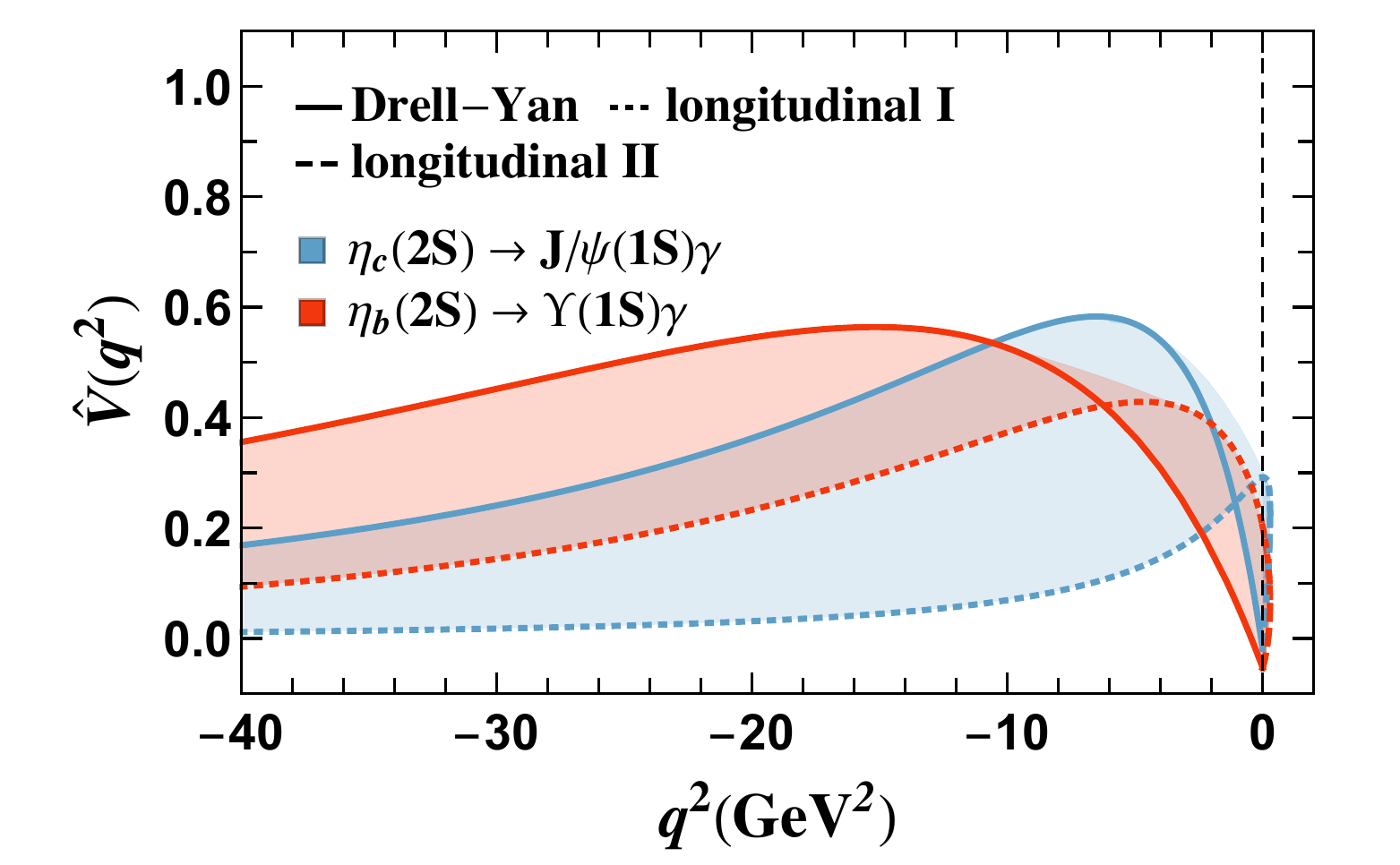}
  \includegraphics[height=6cm]{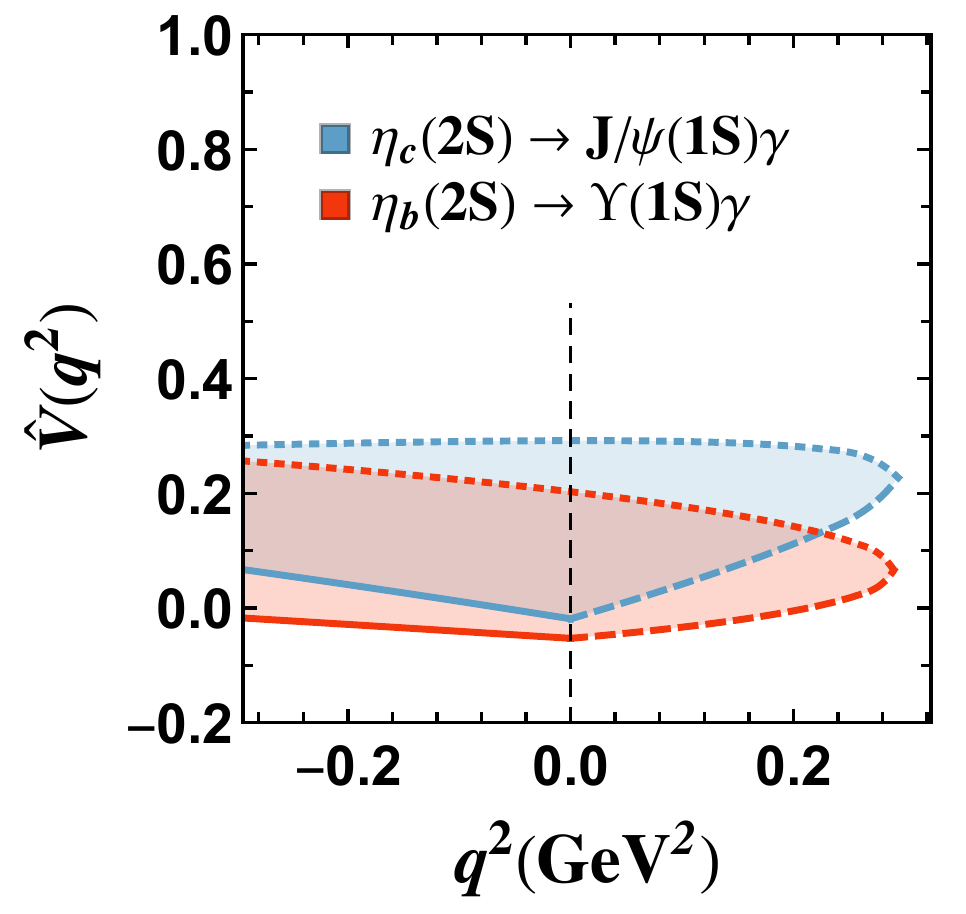}

  \includegraphics[height=6cm]{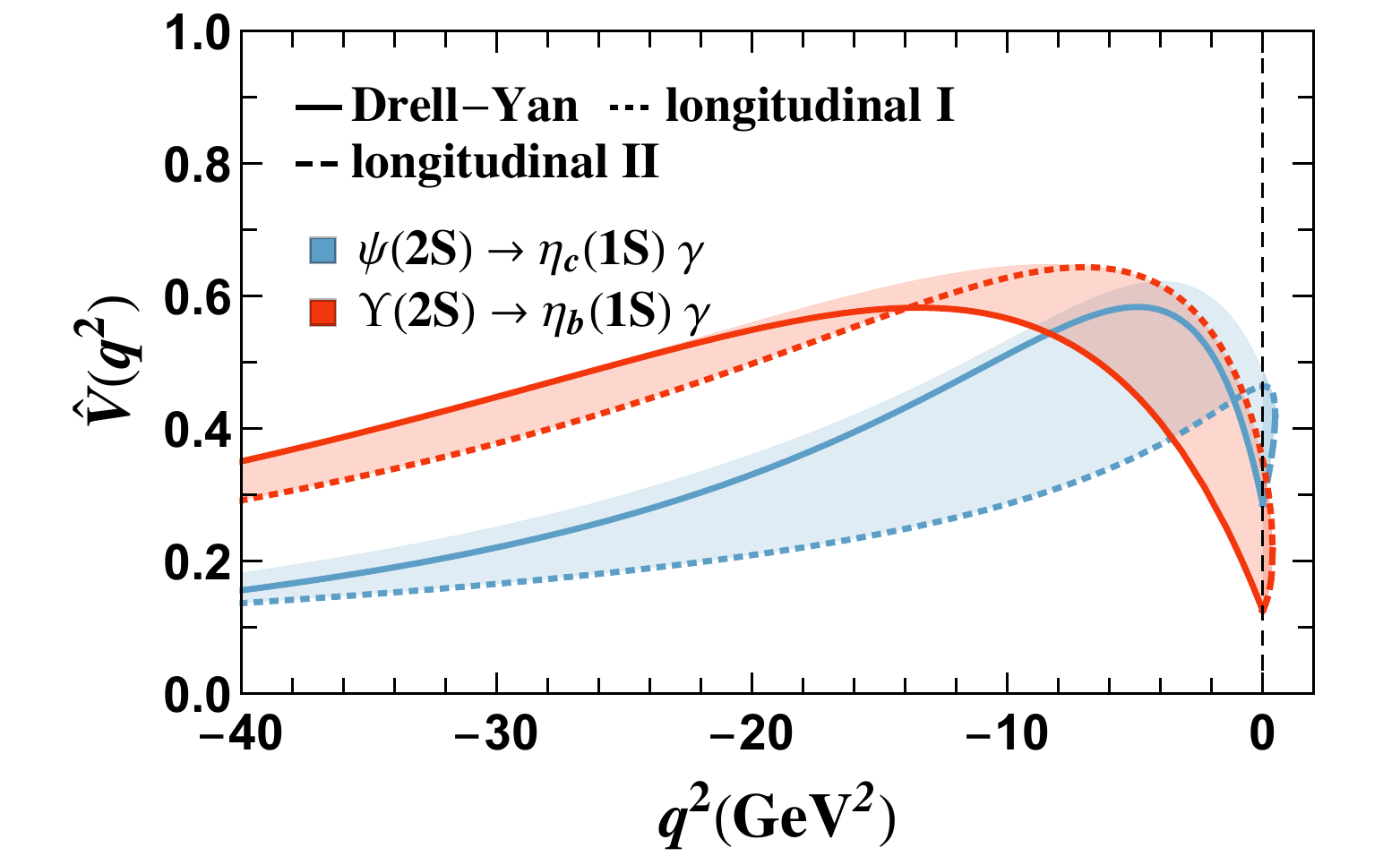}
  \includegraphics[height=6cm]{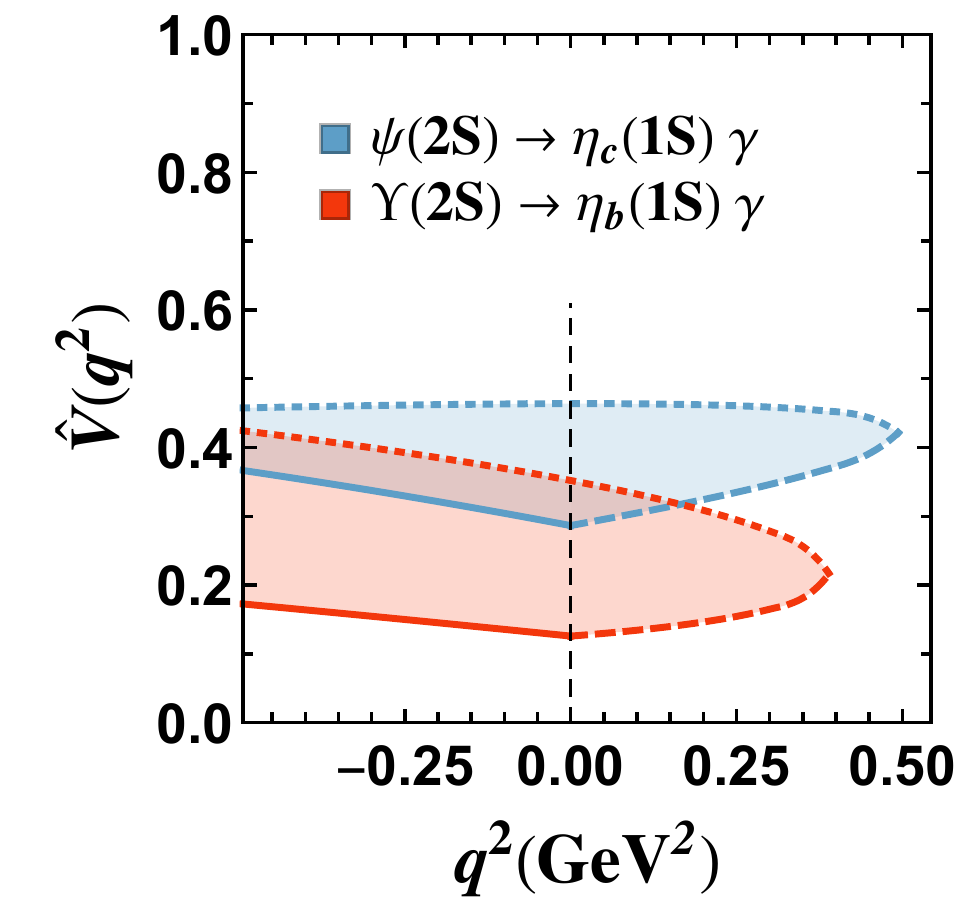}
\caption{The transition form factor of the transition
$\psi_A(2S)\to\psi_B(1S)\gamma$ ($\psi_A,\psi_B= \mathcal V,\mathcal P$ or $\mathcal P, \mathcal V$)
 %$\mathcal{V}(nS)\to\mathcal{P}(n'S)\gamma (n\neq n')$
of charmonia (blue curves/shades) and bottomonia (red curves/shades), calculated with light-front wavefunctions at $N_{\max}=L_{\max}=32$ basis truncation. Meson masses are taken from experimental data~\cite{PDG2018} for defining the frames according to Eq.~\eqref{eq:q2_z_delta}. The solid curves represent the Drell-Yan frame while the other curves represent the longitudinal I (dotted lines) and II (dashed lines) frames. The shaded areas represent the results from all other frames. The left panel shows the transition form factor at a larger scale of $q^2$, and the right panel focuses on the small $q^2$ region. 
%The crosses are the values converted from the corresponding decay widths in PDG~\cite{PDG2018}: $\hat V(0)|_{\psi(2S)\to\eta_c(1S)\gamma}= 0.100(8)$, $\hat V(0)|_{\Upsilon(2S)\to\eta_b(1S)\gamma}=0.070(14)$.
}
\label{fig:TFFccbb2}
\end{figure*}

Figures~\ref{fig:TFFccbb1} and~\ref{fig:TFFccbb2} show numerical results for selected pseudoscalar-vector transition form factors
for charmonia and bottomonia below their respective open-flavor thresholds. 
Those lowest states are the primary focii of several investigations~\cite{PhysRevD.95.074002, PhysRevD.92.094501, Donald_JPsi,Damir2013,PhysRevD.64.074011,bb_GI}. They have been measured in experiments~\cite{PDG2018}, and their transitions are more readily detected with good statistics than higher excited states. Moreover, with their experimental masses, we have an entire landscape of frames in the $(z,\vec\Delta_\perp)$ parameter space according to Eq.~\eqref{eq:q2_z_delta}.
The solid curve represents the Drell-Yan frame, the dotted and the dashed curves represent the two branches of the longitudinal frame, longitudinal I and longitudinal II respectively. The shaded areas represent all other frames with different $z$ and $\Delta_\perp$. We also compare $\hat V(0)$ obtained in different frames with available experimental data from the Particle Data Group (PDG)~\cite{PDG2018} in Table.~\ref{tab:V0}.

\begin{table}[H]
\caption{\label{tab:V0}Comparison of $\hat V(0)$ from available experimental data and the BLFQ calculations in the limiting frames. Values from PDG~\cite{PDG2018} are converted from their decay widths according to Eq.~\eqref{eq:VPwidth}. The BLFQ results are calculated using meson wavefunctions obtained at $N_{\max}=L_{\max}=32$. The Drell-Yan/longitudinal II is the preferred result, and the difference between it and the longitudinal I quantifies the uncertainty resulting from frame dependence. }
\centering
\begin{ruledtabular}
\begin{tabular}{lccc}
\multirow{2}{*}{$\hat V(0)$} & \multirow{2}{*}{PDG~\cite{PDG2018}}
          & \multicolumn{2}{c }{BLFQ} \\
\cline{3-4}
& &Drell-Yan/long-II & long-I\\
\hline
$J/\psi(1S) \to \eta_c(1S)\gamma$ & $1.56(19)$ &2.02  & 2.12 \\
$\eta_c(2S) \to J/\psi(1S)\gamma$ &$\cdots$  &$-0.019$  &0.29 \\
$\psi(2S) \to \eta_c(1S)\gamma$ & $0.100(8)$ &0.29  & 0.46 \\
$\psi(2S) \to \eta_c(2S)\gamma$ & $2.52(91)$ &2.09  & 2.14 \\
$\Upsilon(1S) \to \eta_b(1S)\gamma$ &$\cdots$  &2.01  & 2.03 \\
$\eta_b(2S) \to \Upsilon(1S) \gamma$ &$\cdots$  &$-0.052$  &0.20\\
$\Upsilon(2S) \to \eta_b(1S)\gamma$ & $0.070(14)$ &0.13  & 0.35 \\
$\Upsilon(2S) \to \eta_b(2S)\gamma$ &$\cdots$ & 2.02 & 2.03 \\
\end{tabular}
\end{ruledtabular}
\end{table}

For the transition form factor of the \emph{allowed} transition, i.e. $\psi_A(nS)\to\psi_B(nS)\gamma$, $(\psi_A, \psi_B = \mathcal{V}, \mathcal{P} \text{ or }\mathcal{P}, \mathcal{V})$, as in Fig.~\ref{fig:TFFccbb1}, there are no crossings between the curves of the longitudinal frame and the Drell-Yan frame. In these cases, the results from all other frames are represented by the enclosed shaded area. The frame dependence is relatively small, no more than a $5\%$ spread at $q^2=0$, as in Table.~\ref{tab:V0}. For the transition form factor of the \emph{hindered} transitions, i.e. $\psi_A(nS)\to\psi_B(n'S)\gamma (n'\neq n)$, $(\psi_A, \psi_B = \mathcal{V}, \mathcal{P} \text{ or }\mathcal{P}, \mathcal{V})$, as in Fig.~\ref{fig:TFFccbb2}, the curves of the longitudinal frame and the Drell-Yan frame cross each other, and their joined lower bound forms the lower bound for the results from all other frames. The upper bound, however,  envelops the Drell-Yan and the longitudinal results. The frame dependence of these hindered transitions is very strong, indicating major sensitivity to the Lorentz symmetry breaking. This sensitivity seems understandable since these weaker transitions result from cancellations coming from different regions of integration.
%is also comparatively large, could get to as much as $200\%$ at $q^2=0$.

We also compare charmonia and bottomonia at corresponding transition modes in Figs.~\ref{fig:TFFccbb1} and ~\ref{fig:TFFccbb2}. Such comparisons suggest that the frame dependence tends to be reduced for heavier systems, presumably due to the overall reduction in relativistic effects.

%({\color{red}Yang: In my opinion, the part with $|q^2| > max\{ \kappa^2*N_{max}, L_{max}*M^2 \}$ should not be shown, since they are dominated by the basis asymptotes and are of no physical interests.}
%{\color{blue} The largest supported $q^2$ would be $31~\GeV^2 (44~\GeV^2)$ for charmonia (bottomonia). In Figs.4-5, I intend to show that the transition form factor approaches 0 at large $q^2$, so I plotted them down to $-60~\GeV^2$. But in Fig.6, which is about basis sensitivity, I used the supported range. }
%{\color{red}Yang: This is however very misleading. The readers may think our calculation is valid up to $60 \GeV^2$. Besides, the fact that the form factor approaches to zero is a basis effect at the scale you are showing.}
%{\color{cyan} Ok, I changed the range to $40 \GeV^2$. I see that you plotted the elastic form factors up to $60 \GeV^2$ and even $80 \GeV^2$ in the frame dependence paper. Do they have a different supported range? } )

\begin{figure*}[ht]
  \centering
  \includegraphics[width=0.48\textwidth]{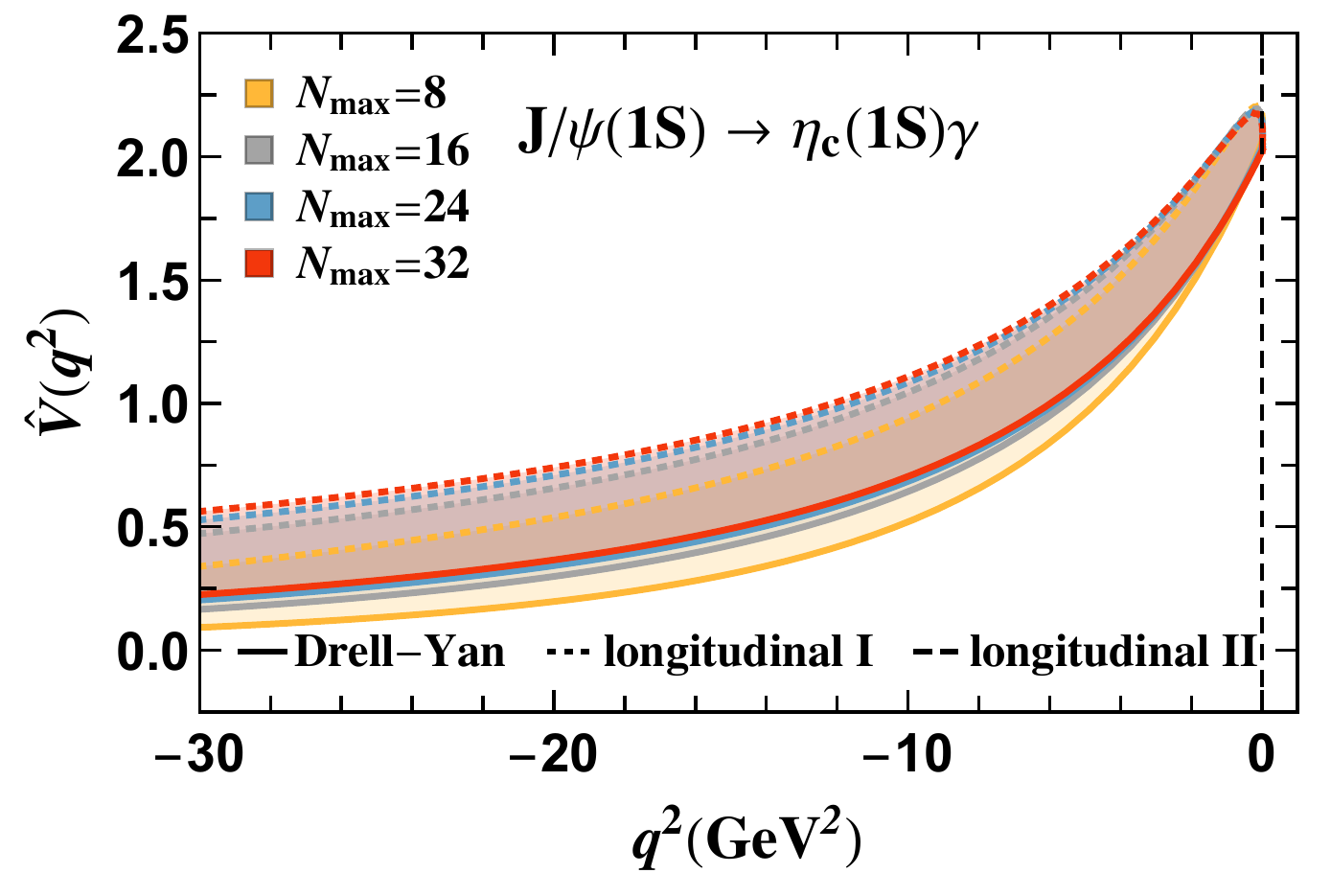}
  \includegraphics[width=0.48\textwidth]{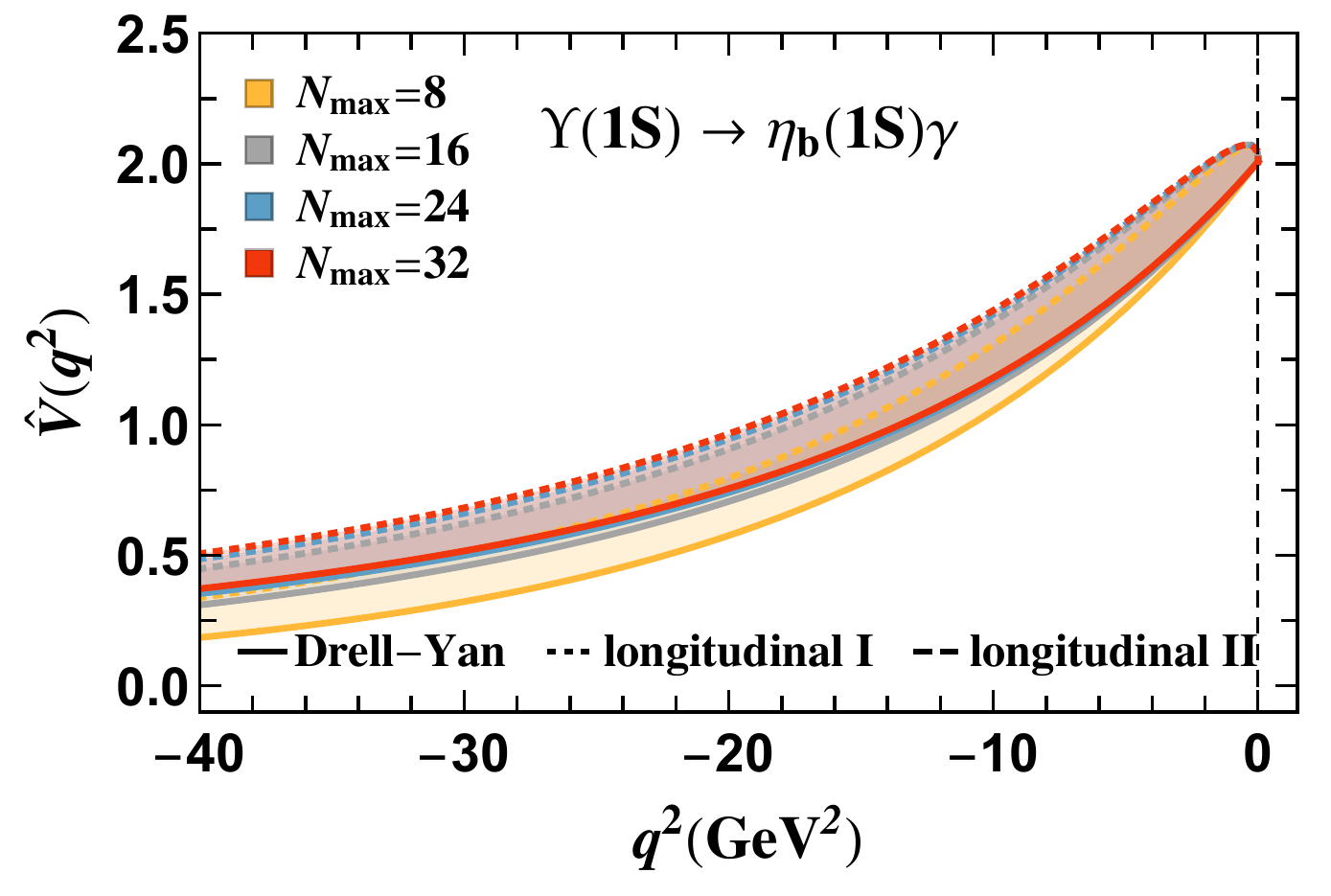}

  \includegraphics[width=0.48\textwidth]{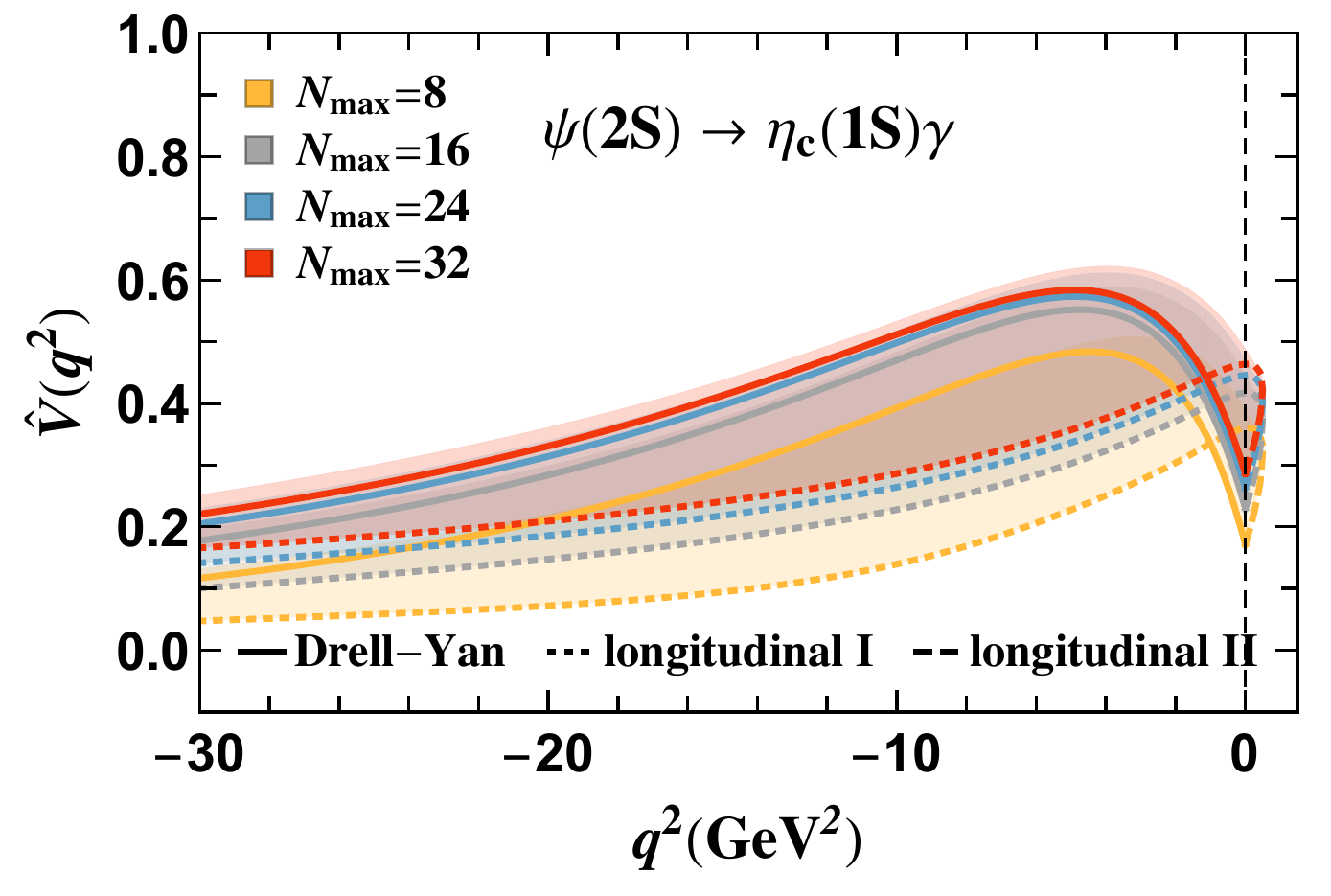}
  \includegraphics[width=0.48\textwidth]{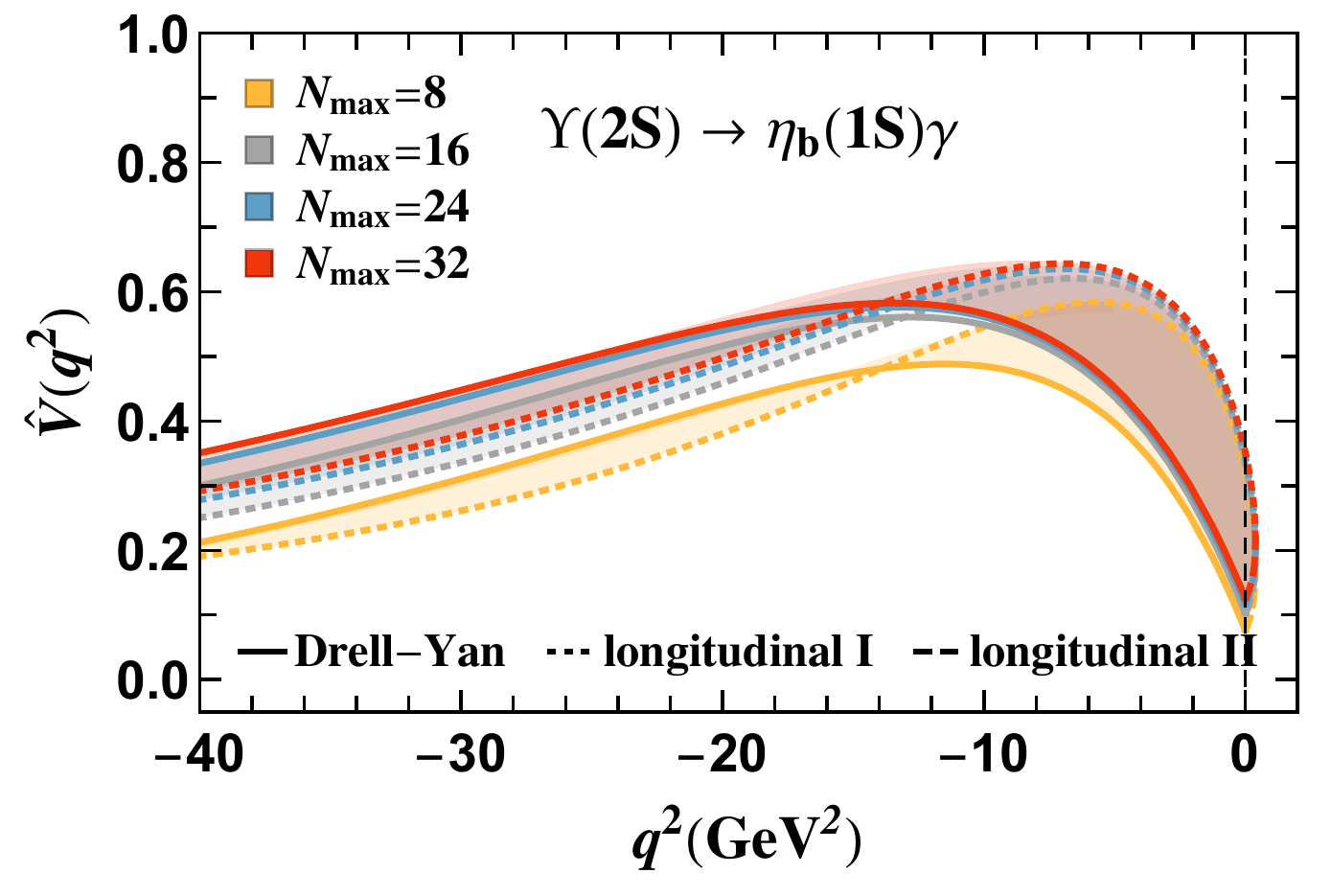}

  \caption{The transition form factors for charmonia (left panels) and bottomonia (right panels) with different basis truncations. Meson masses are taken from experimental data~\cite{PDG2018} in defining the frames according to Eq.~\eqref{eq:q2_z_delta}. The solid curves represent the Drell-Yan frame while the other curves represent the longitudinal I (dotted lines) and II (dashed lines) frames. The shaded areas represent the results from all other frames.}
 \label{fig:TFF_Nmax}
\end{figure*}

It is natural to ask how frame dependence may be sensitive to the BLFQ basis truncation applied to these valence Fock space calculations. For this purpose, we present transition form factors from different basis truncations in Fig.~\ref{fig:TFF_Nmax}. A trend towards convergence with increasing basis cutoff is observed in both the Drell-Yan and the longitudinal frames. The frame dependence indicated by the shaded regions is shrinking slightly with increasing basis cutoff but Lorentz symmetry breaking effects remain visible even at the highest basis cutoffs.

From those results, we observe that the frame dependence  of the transition form factor can be characterized by the two limits,
the Drell-Yan and the longitudinal frames. Transitions with excitations in the lighter system (e.g. $\eta_c(2S)\to
J/\psi(1S)\gamma$) admit the largest frame dependence, implying a stronger sensitivity to the Fock sector truncation. Our
suggested frames for the calculation in the valence Fock sector, the Drell-Yan and the longitudinal-II frames, provide
values of $\hat V(0)$ that are closer to the experimental data, as seen in Table.~\ref{tab:V0}.

\subsection{The electromagnetic Dalitz decay}\label{subsec:Dalitz}
\begin{figure*}[htp!]
  \centering
  \includegraphics[width=0.48\textwidth]{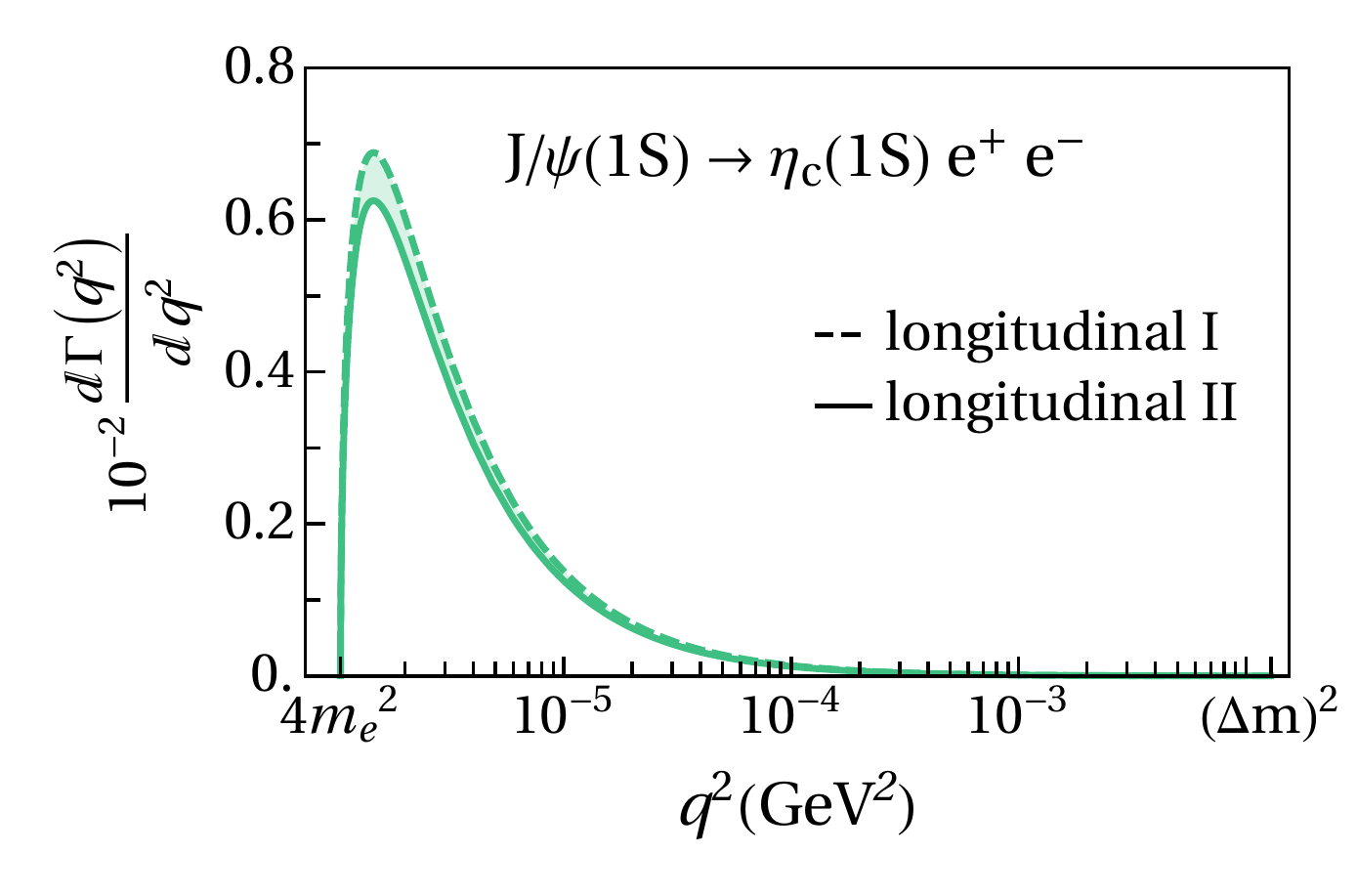}
  \includegraphics[width=0.48\textwidth]{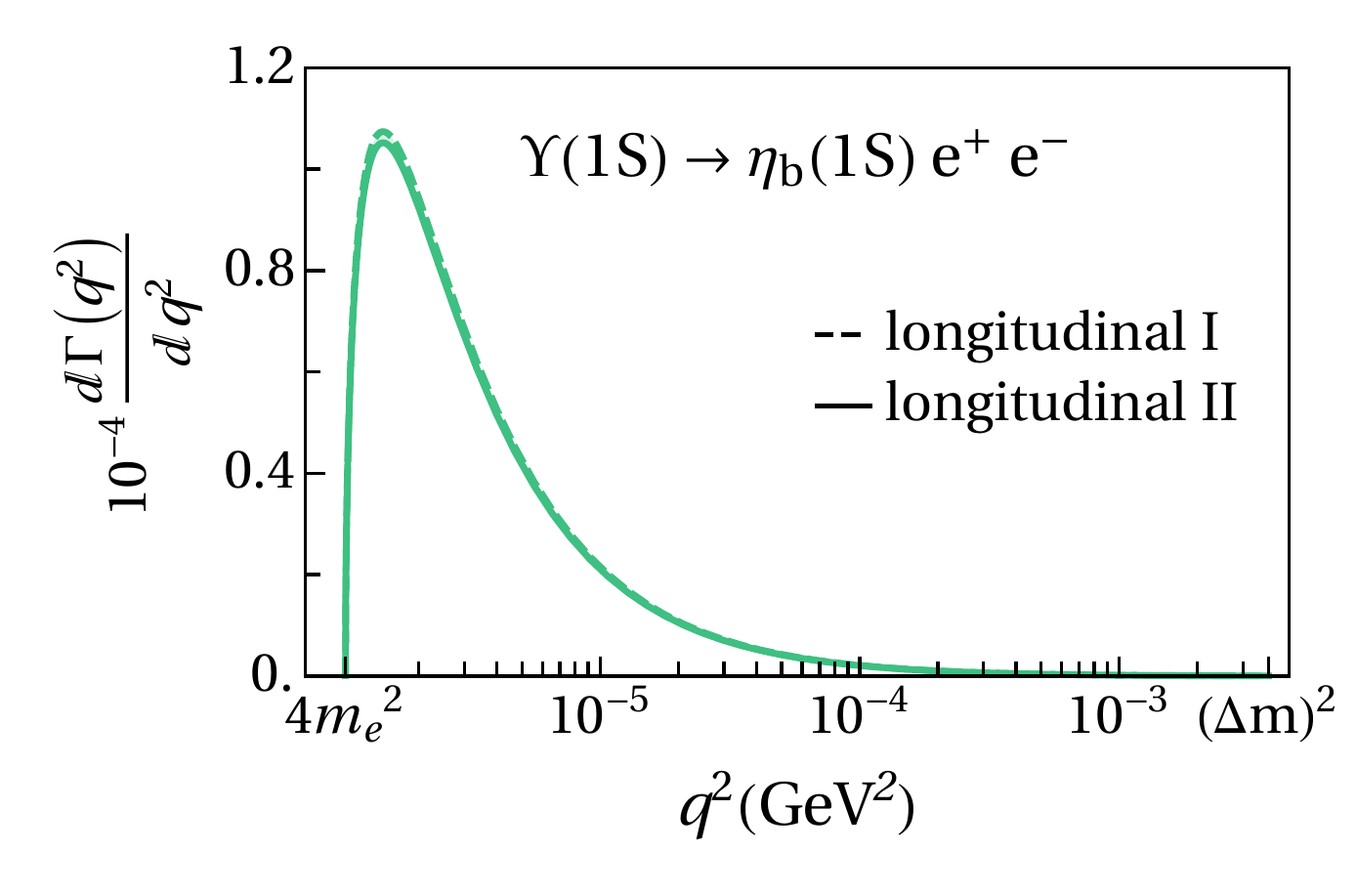}

  \includegraphics[width=0.48\textwidth]{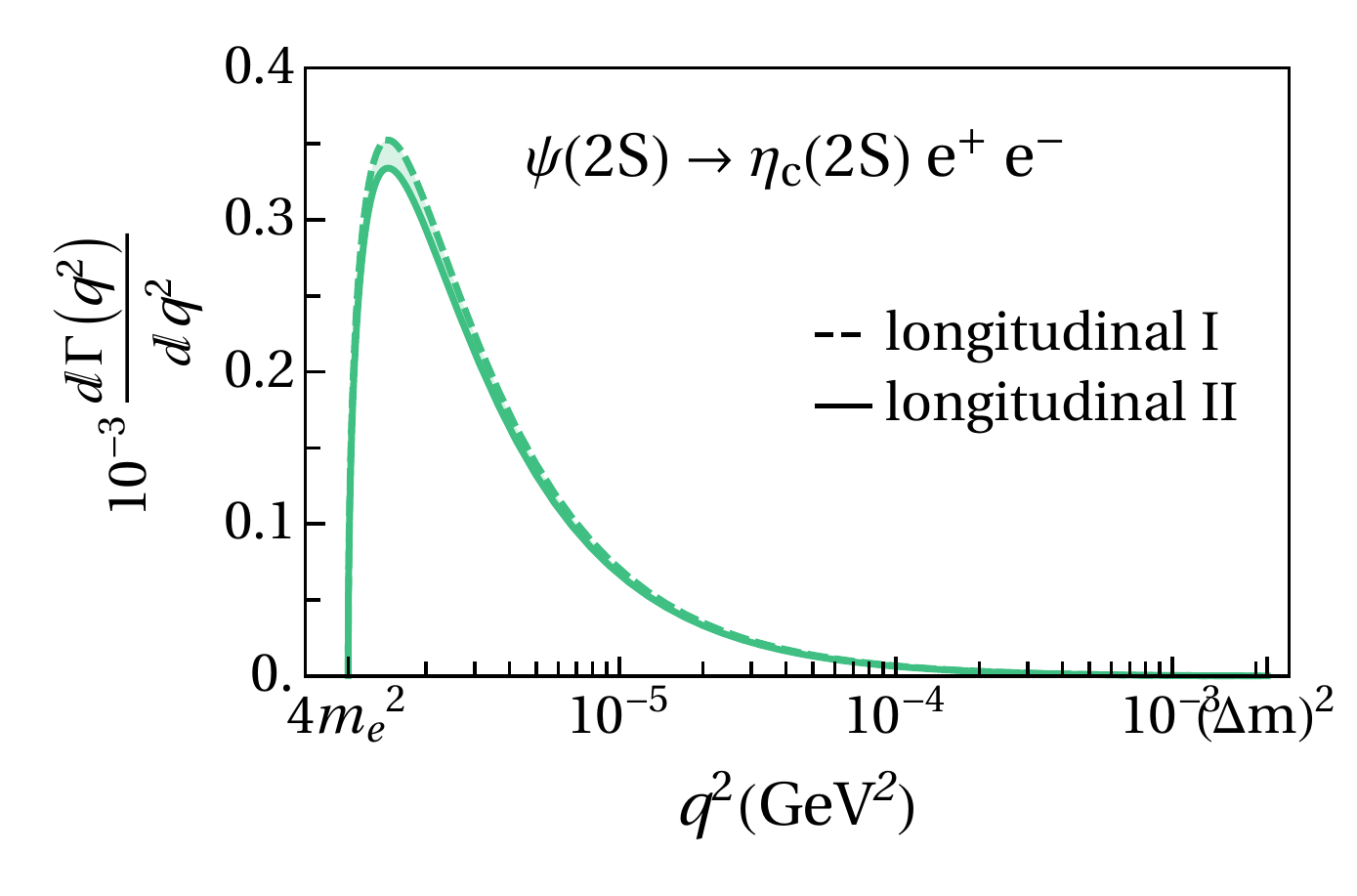}
  \includegraphics[width=0.48\textwidth]{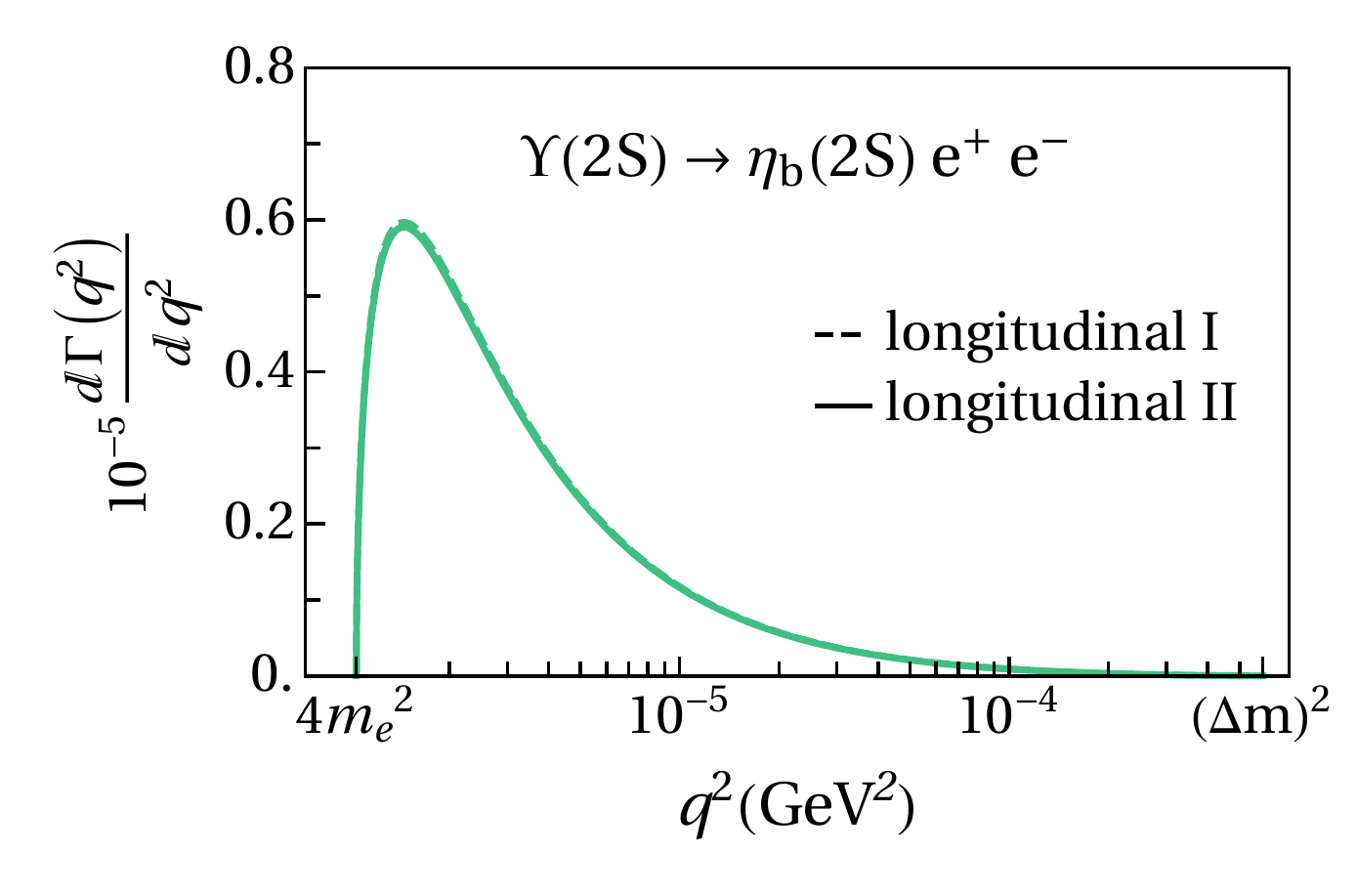}

  \includegraphics[width=0.48\textwidth]{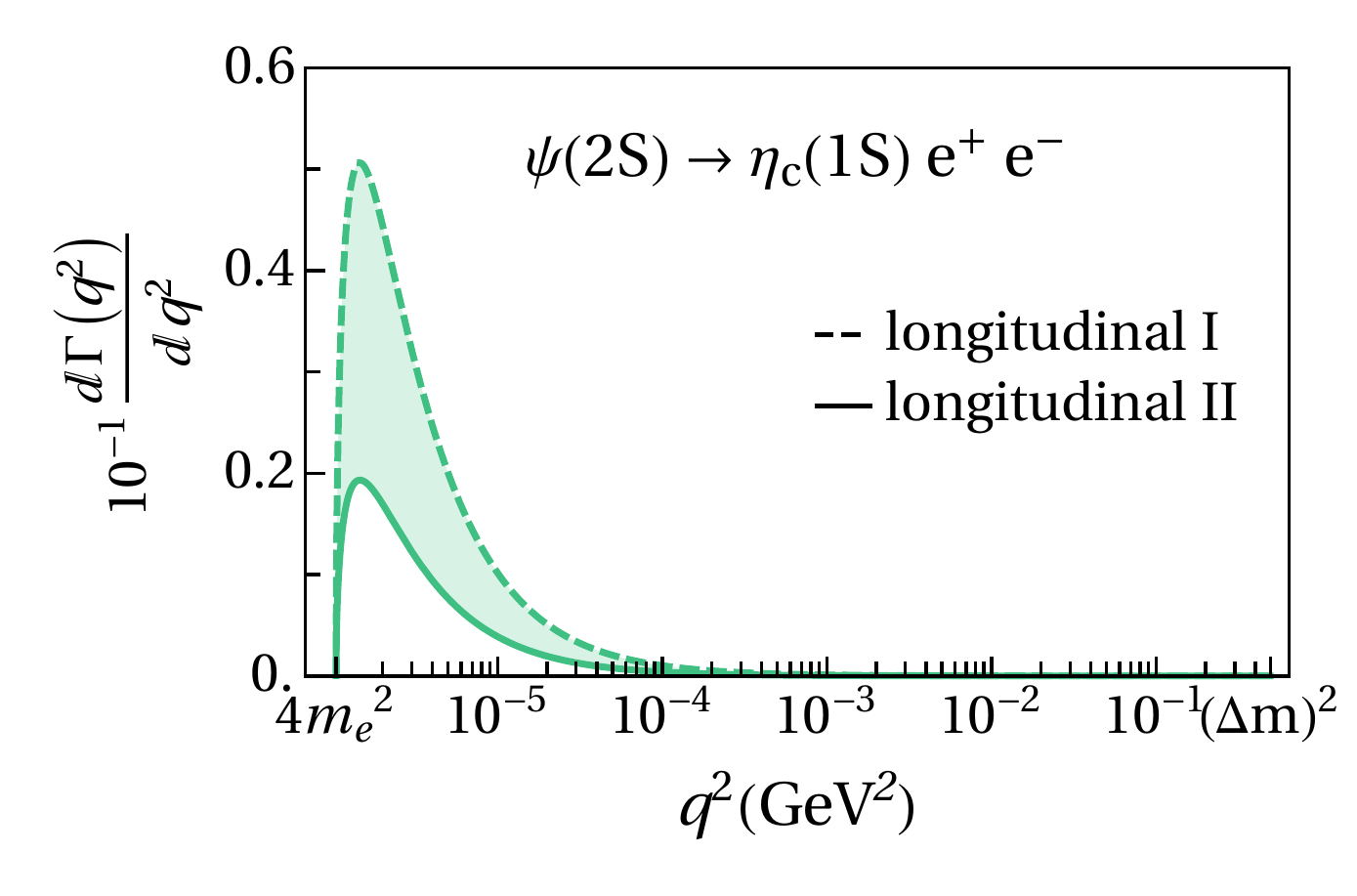}
  \includegraphics[width=0.48\textwidth]{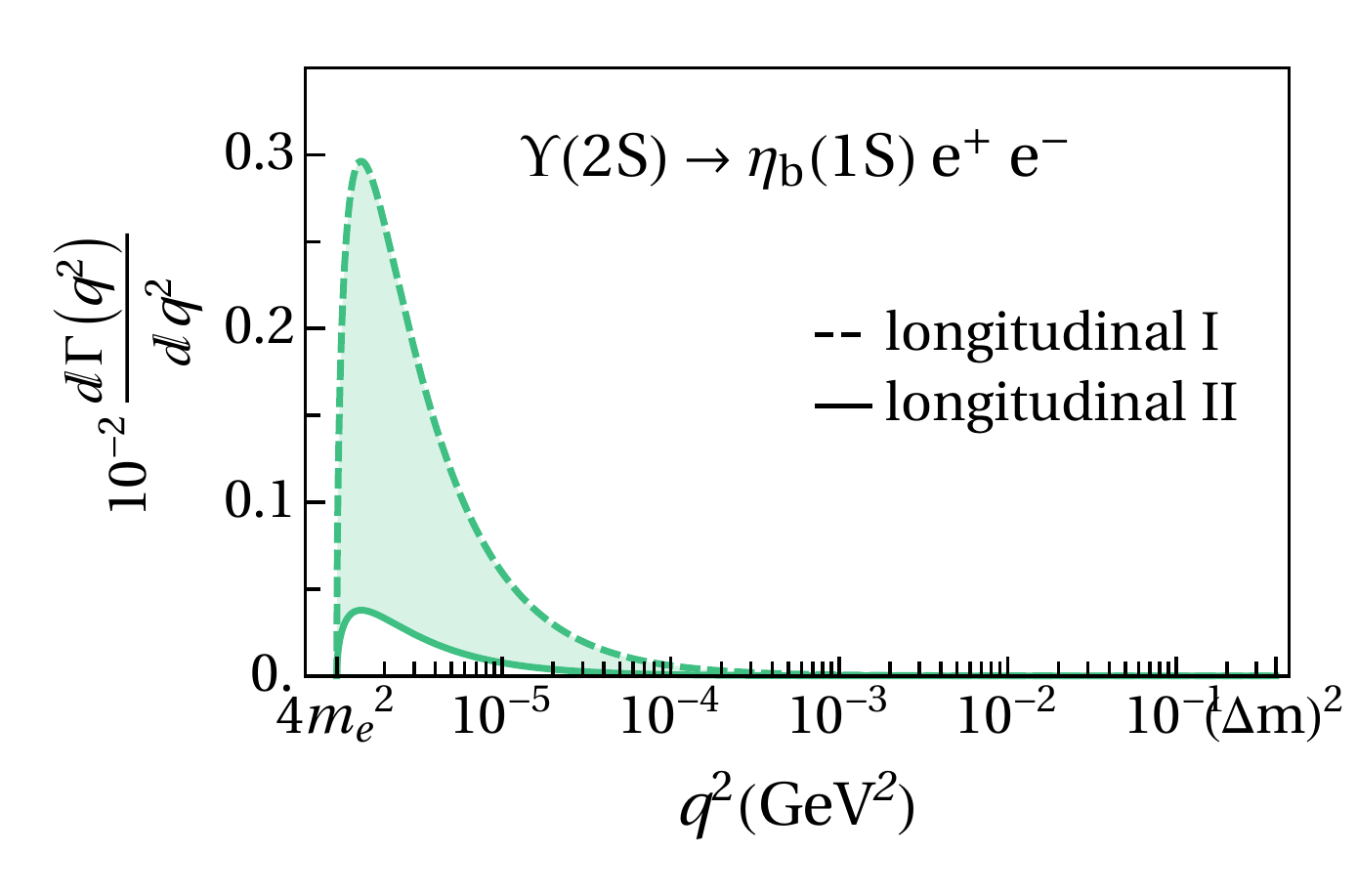}

  \includegraphics[width=0.48\textwidth]{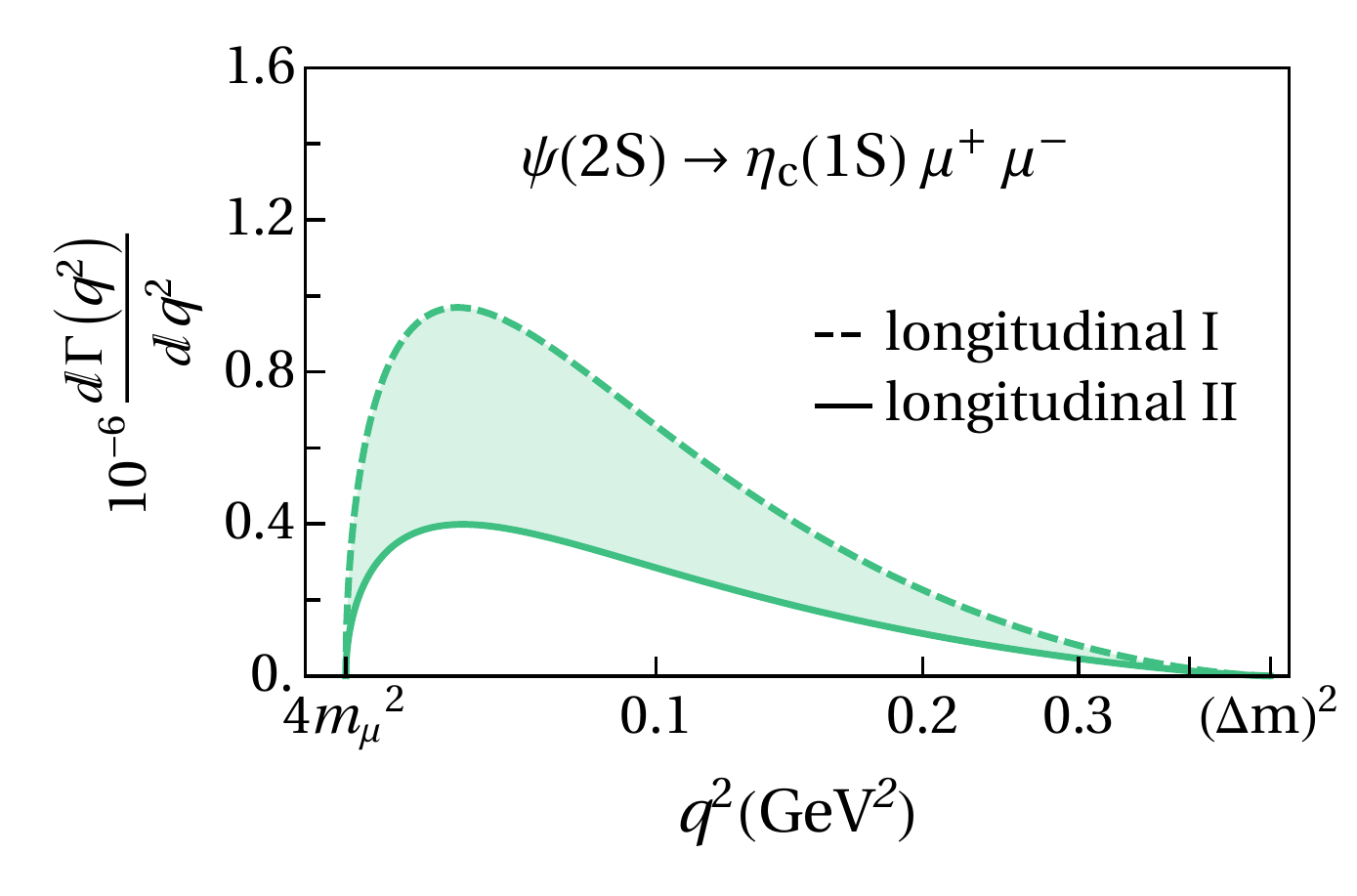}
  \includegraphics[width=0.48\textwidth]{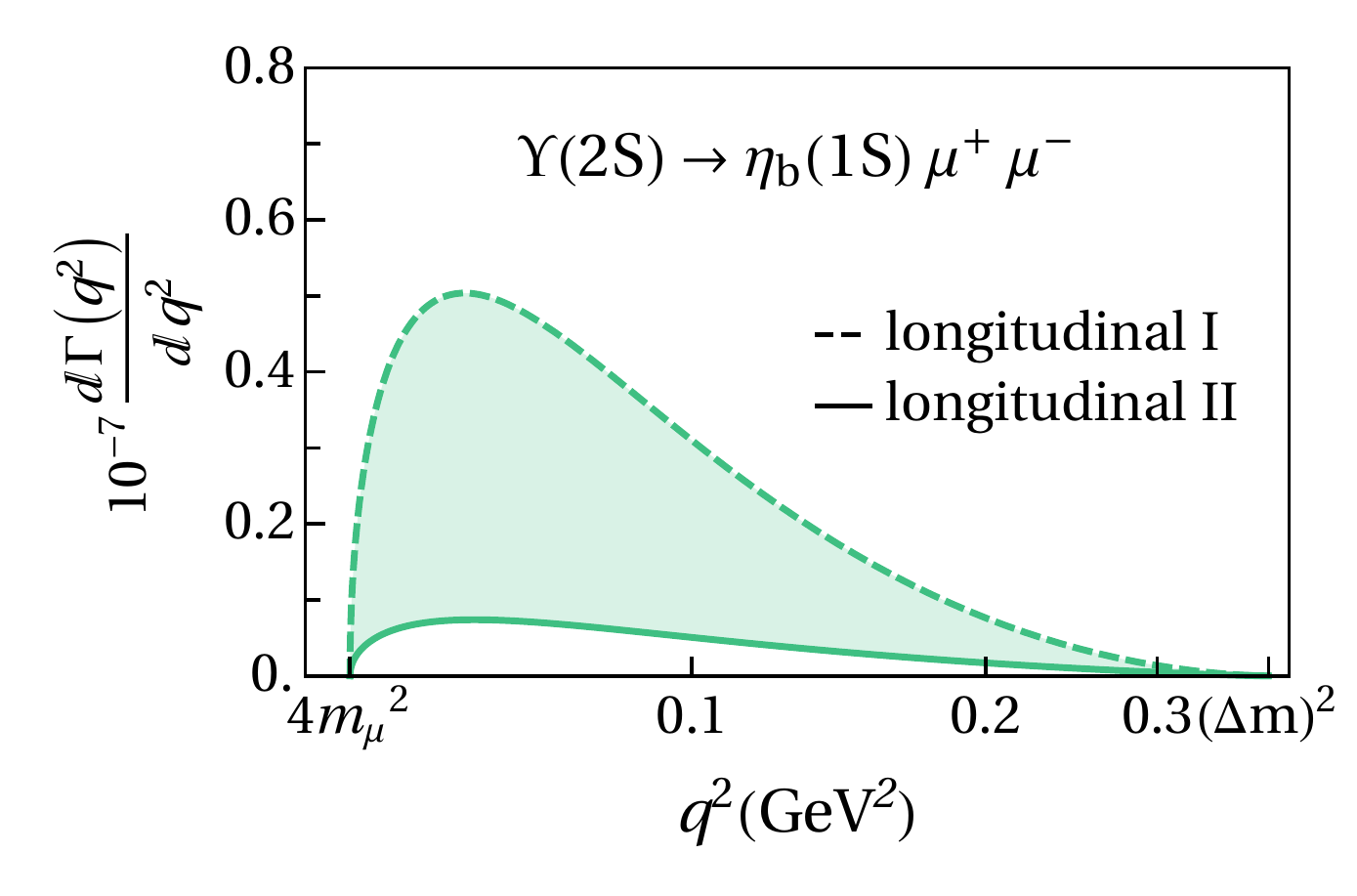}

  \caption{The effective mass spectrum of the lepton pairs in the Dalitz decays for charmonia (left panels) and bottomonia (right panels). The dashed and solid curves represent the longitudinal I and II frames respectively. The shaded areas represent the results from all other frames. $\Delta m^2 = (m_A - m_B)^2$ is the square of the mass difference between the initial and the final mesons. Meson masses are taken from experimental data~\cite{PDG2018}. }
 \label{fig:Dalitz}
\end{figure*}

The effective mass spectrum of the lepton pair in the Dalitz decay can be obtained from the corresponding transition form factor according to Eqs.~\eqref{eq:VPwidth} and~\eqref{eq:D_width}. The results of $\diff \Gamma(\psi_A \to \psi_B l^+ l^-)/\diff q^2 $ for eight selected decays 
%({\color{red}Y.Li: Could we show more of these results? Also, are there any experimental data?}
%{\color{blue}M.Li The relation of $q^2$ to $z,\vec\Delta_\perp$ requires the knowledge of the meson masses. As a result, if a meson state is not measured by experiment, the frames in terms of $z,\vec\Delta_\perp$ would not be well defined. For this reason, I did not present results of transitions including $\eta_b(3S), \Upsilon(1D), \Upsilon(2D)$ states. There are no experimental data for those transitions yet.}
%{\color{red}Y.Li: One of the possibilities is to use the calculated masses.}
%{\color{blue}M.Li: Using calculated masses for unmeasured states and using the experimental masses for measured states would cause inconsistency. The BLFQ $\eta_b(3S)$ mass is larger than the experimental $\Upsilon(3S)$ mass. Since the decay width, and thus the converted experimental transition form factor, is very sensitive to the meson masses, we use the experimental masses.} 
%{\color{red}Y.Li: Does the form factor V depend on the masses? I thought it only depends on wave functions. In any case, there are some ways to present more predictions. }
%{\color{blue}M.Li: The form factor V does not depend on the meson masses, but the definition of the frames depends on the meson masses, see Eq.(4). To calculate the transition form factor V at a certain $q^2$, we need the values of $z$ and $\vec\Delta$, thus the meson masses. In the Drell-Yan frame ($z=0$), $q^2=-\vec \Delta_\perp^2$, the meson masses are not involved. But in general frames, we need to know the meson masses.})
 are shown in Fig.~\ref{fig:Dalitz}. The frame dependence is barely visible in the allowed transitions as in the top four panels, but very substantial in the hindered transitions as in the bottom four panels. Such different sensitivities to frames can be expected in light of the sensitivities observed for the transition form factors in the time-like region. The allowed transitions are between states with similar spatial wavefunctions (e.g. $J/\psi(1S)\to \eta_c(1S)e^+ e^-$), whereas the hindered transitions are between states with nearly orthogonal spatial parts (e.g. $\psi(2S)\to \eta_c(1S)e^+ e^-$). Therefore in the latter cases, the transition form factors and thus the leptonic widths would admit strong cancellations between positive and negative contributions, and thus become more sensitive to the finer details of light-front wavefunctions. %We know that, in different frames, the transition is actually probing the wavefunctions differently, as in Eq.~\eqref{eq:Vmj0_red}. In consequence, the hindered transitions are more sensitive to the frames.}
\section{Summary and outlook}\label{sec:summary}
We have introduced the M1 transition form factor on the light front in a general frame. We analyzed the contributions from the parton-number-conserving
term ($n\to n$) and the non-conserving term  ($n+2\to n$) in different frames, and suggested that frames with minimal $z$ values
could suppress the latter. Therefore, the minimal-$z$ frames, i.e. the Drell-Yan and the longitudinal-II frames, are preferred for calculating
transition form factors in the valence Fock sector.

We then looked into the heavy quarkonia
system and showed that, using light-front wavefunctions from the valence Fock sector, the transition form factor admits moderate
frame dependence in the case of allowed transitions. For the case of hindered transitions, we find that the frame dependence is more severe. Our results from different frames fall between the two special frames, the Drell-Yan and the
longitudinal frames, a pattern also observed in the study of elastic form factors~\cite{Li:2017uug}. In the space-like region, the
Drell-Yan and the longitudinal-I frames form two limits, while in the time-like region, the limits become the two branches of
the longitudinal frame (referred as longitudinal I and II). We employed the difference of results in the Drell-Yan and the longitudinal frames as a metric for the violation of the Lorentz symmetry due to Fock space truncation. 

With the transition form factor in the time-like region, we obtained the decay widths of the associated Dalitz decay and use the
frame difference as an uncertainty for the result. We hope that the comparison of our predictions with future experiments could
help justify our choice of the preferred frames for valence Fock sector calculations. By exhibiting frame dependence in these form factors and decay widths, we motivate solving quarkonia systems in higher Fock sectors ~\cite{Yukawa4_Li:2015iaw, FSDR_Karmanov:2016yzu} and providing a more complete description with increasingly precise treatments of the Lorentz symmetry.

We expect that by considering contributions from higher Fock sectors, this frame dependence would eventually vanish. As we have discussed in Sec.~\ref{sec:TFF}, in terms of the transition form factor, the contribution from the valence sector and that from the higher sectors are different in different frames, but the full result (the summation of the two) should be invariant. The comparison of the future full transition form factor to the valence result in different frames, as presented here, could verify our deduction on the suggested frames. In practical calculations, if taking into account the higher Fock sector contribution does not resolve all the frame dependence, the residue is likely caused by approximations in modeling the Hamiltonian and numerical uncertainties. Checking the frame dependence could still provide a measurement 
on the violation of the Lorentz symmetry for the phenomenological model of the system.

It should be noted that though we take heavy quarkonia as a concrete study object in this paper, the formalism of the transition form factors in different frames also applies to light mesons and could be extended to other systems such as baryons.
%Another hope of including higher Fock sectors is to study the dependence of the transition form factors on the current components. Like the frame dependence, current component dependence also reveals the violation of Lorentz covariance in light-front dynamics. In this work, we have inherited the choice of the transverse current from our previous investigation~\cite{Li:2018uif}. 

\section*{Acknowledgments}
We wish to thank Shaoyang Jia, Tobias Frederico, Shuo Tang, Wenyang Qian, Anji Yu, Xingbo Zhao, Ho-Meoyng Choi and Chueng-Ryong Ji for valuable discussions.
This work was supported in part by the US Department of Energy (DOE) under Grant Nos. DE-FG02-87ER40371, DE-SC0018223 (SciDAC-4/NUCLEI), DE-SC0015376 (DOE Topical Collaboration in Nuclear Theory for Double-Beta Decay and Fundamental Symmetries). This research used resources of the National Energy Research Scientific Computing Center (NERSC), a U.S. Department of Energy Office of Science User Facility operated under Contract No. DE-AC02-05CH11231.

\appendix

\bibliography{TFF_Dalitz}
\end{document}